\documentclass[fleqn,10pt]{wlscirep}
\usepackage{CJKutf8}
\usepackage[T1]{fontenc}
\usepackage{graphicx}  

\usepackage{bm}
\usepackage{amssymb}
\usepackage{amsmath}
\usepackage{float}

\newcommand{\ham}{\hat{\mathcal{H}}}
\newcommand{\dg}{\dagger}
\newcommand{\w}{\omega}

\title{Cavity-Magnon-Polariton spectroscopy of strongly hybridized electro-nuclear spin excitations in LiHoF$_4$}
\author[1,2]{Yikai Yang}
\author[2]{Peter Babkevich} 
\author[2]{Richard Gaal}
\author[2]{Ivica {\v Z}ivkovi{\'c}}
\author[2,*]{Henrik M. R{\o}nnow}
\affil[1]{Department of Engineering Science, University of Oxford, Parks Road, Oxford OX1 3PJ, United Kingdom}
\affil[2]{Laboratory of Quantum Magnetism. Institute of Physics, École polytechnique fédérale de Lausanne, Lausanne CH-1015, Switzerland}
\affil[*]{henrik.ronnow@epfl.ch}
\date{\today}

\begin{abstract}
We first present a formalism that incorporates the input-output formalism and the linear response theory to employ cavity-magnon-polariton coupling as a spectroscopic tool for investigating strongly hybridized electro-nuclear spin excitations. A microscopic relation between the generalized susceptibility and the scattering parameter $|S_{11}|$ in strongly hybridized cavity-magnon-polariton systems has been derived without resorting to semi-classical approximations. The formalism is then applied to both analyze and simulate a specific systems comprising a model quantum Ising magnet (LiHoF$_4$) and a high-finesse 3D re-entrant cavity resonator. Quantitative information on the electro-nuclear spin states in LiHoF$_4$ is extracted, and the experimental observations across a broad parameter range were numerically reproduced, including an external magnetic field traversing a quantum critical point. The method potentially opens a new avenue not only for further studies on the quantum phase transition in LiHoF$_4$ but also for a wide range of complex magnetic systems.
\end{abstract}

\begin{document}
\begin{CJK*}{UTF8}{gbsn}

\flushbottom
\maketitle

\section*{\label{sec:Intro} Introduction}
The light-matter interaction has long been a focal point of intense research \cite{Dicke1954,Lee1989,Zakowicz1974,Tavis1967}. It is not only a fundamental aspect of our universe but also crucial for studying of a wide variety of condensed matter materials. Furthermore, it is pivotal for candidate systems for quantum communication and quantum computing as it is the most promising means to facilitate qubit manipulations \cite{Cox2019} as well as information transmission among quantum computing elements \cite{Kaiser2019}. Most existing studies have concentrated on manipulating ensembles of weakly interacting spins typically found in either optical traps or a paramagnetic crystals \cite{Abe2011,Cox2019,Herskind2009,Schuster2010}. Only recently has research on manipulating spins in strongly interacting magnetic compounds using light begun to gain traction, with much of the focus being on the single Kittel mode in yttrium iron garnet (YIG) \cite{McKenzie2019,Flower2019,Zhang2017}. 

Traditionally, spin excitation spectra are measured using techniques such as electron spin resonance (ESR) and nuclear magnetic resonance (NMR), which are built upon very similar principles but target different processes in a magnetic sample. While both techniques are relatively mature in their own individual domain, strong hyperfine interaction causes complications for either application. On the one hand, it would likely move the interested spin transitions out of the frequency range of conventional NMR techniques and at the same time greatly shorten the spin coherence time, rendering the technique insensitive to the dynamics. On the other hand, conventional ESR spectroscopy by design directly addresses electronic spin transitions, and a strong hyperfine interaction incurs considerable broadening of line widths of these transitions. Inelastic neutron scattering is another frequently employed method when investigating spin excitations, it has the advantage of being able to measure crystal momentum dependent excitations across a wide energy scale. However, it is limited by its energy resolution \cite{Ehlers2009}. As a result of these difficulties, to date, measurements on samples with multi-mode spin excitations and strong hyperfine interactions have been scarce.

In the present study, we exploit the strong hyperfine interaction to indirectly measure and manipulate nuclear spin states through interfacing with the electronic spins, in other words, an nuclear magnetic resonance (NMR) measurement facilitated by electron spin resonance (ESR) measurement. The method is applied to LiHoF$_4$, an insulating rare-earth magnet with large electronic and nuclear spins (J = 8, I = 7/2) that has been established as a model 3D Ising magnet undergoing a quantum phase transition (QPT) induced by transverse magnetic field \cite{Bitko1996,Chakraborty2004,Ronnow2005,Ronnow2007}. The localized hyperfine interaction on Ho$^{3+}$ sites in LiHoF$_4$ is very strong, which fundamentally alters the collective energy landscape and is responsible for the incomplete softening of the electronic mode \cite{Ronnow2005,McKenzie2018} at the quantum critical point (QCP). Concomitantly, it results in field-dependent excitation spectra in the GHz frequency range with long life-times \cite{Kovacevic2016, Libersky2021}, providing the compound potentials for applications in quantum communications such as spatial multiplexing.

In order to perform the "ESR assisted NMR" measurement described above on LiHoF$_4$, we strongly couple a single crystal sample of LiHoF$_4$ to the photon field inside a re-entrant cavity, thereby creating a cavity-magnon-polariton (CMP) system. Such systems are traditionally modeled as two- or multi-level-systems (TLS/MLS) interacting with a bosonic bath that represents the cavity field. These systems are often characterized within the framework of the input-output formalism with frequent use of either macrospin representation \cite{Soykal2010} or Bosonization procedures \cite{Zhang2014} for simplification. While this provides some technical convenience when dealing with a single magnon mode, it quickly becomes cumbersome as the number of transitions increases, which is the case for most real materials. Furthermore, in deriving expressions for the scattering parameters under these procedures, rotating wave approximation (RWA) is often applied to the interaction between the TLS/MLS and the cavity field to simplify the calculation, limiting the applicability of the theory. 

In contrast, we start with a total Hamiltonian using reaction coordinate and circumvent both semi-classical approximations by establishing an explicit connection between the input-output formalism and the generalized susceptibilities from the linear response theory. We then apply this model to analyze our experimental data on LiHoF$_4$ in a cavity at the radio frequency (RF), followed by numerically reproducing all observed behaviors to demonstrate the model's capacity as both an investigative instrument for magnon-polariton systems alike as well as a benchmark for theories that aim to provide comprehensive descriptions of spin dynamics in real materials \cite{Everts2019}.

\section*{\label{sec:data} Results \& Discussion}
\subsection*{\label{apx:iptopt}Theoretical Results}
\subsection*{Scattering Parameter of a Cavity-Magnon-Polariton System}
The study of the hybridization between an ensemble of weakly interacting spins and a single-mode-cavity field has long been a subject of intense research \cite{Tavis1968,McKenzie2019,Lee1989,Greiner2001,Scully1997}. A significant milestone was the establishment of the input-output formalism established in late 1980s \cite{Gardiner1985}, which has since been widely applied in the field of quantum optics and has consistently proven its efficacy \cite{Walls2008, Meystre2007,Garrison2008}. The formalism has been instrumental in the development of cavity-magnon-polariton (CMP) systems \cite{Fan2010,Xu2015,Li2018}. In the present study, we begin by adopting the same approach in treating the coupling between an spin ensemble and the cavity electromagnetic (EM) field as a magnon-polariton system. We then derive a microscopic relation between the scattering parameter $|S_{11}|$ and the magnetic susceptibility of the sample, which eventually leads to a formula that can adequately describe such a complex system across a wide range in the parameter space. 

To start, a cavity magnon-polariton system can be conveniently broken down to three parts: 1) the Cavity resonator. 2) A spin ensemble and 3) the environment (reservoir) outside the cavity.  The system's total Hamiltonian can therefore be written as:
\begin{equation}
\ham = \ham_{cavity} + \ham_{reservoir} + \ham_{spin} + \ham^{int}_{cav-res} + \ham^{int}_{cav-spin} + \ham^{int}_{res-spin}
\end{equation}
where the first three terms represent the diagonal Hamiltonians of the three individual subsystems, whose pairwise interactions are captured by the last three terms. At this stage, the specific form of $\ham_{spin}$ is irrelevant; it is only assumed to be hermitian and therefore can be unitarily diagonalized. It then follows that there exists a complete set of eigenstates $|n\rangle$ so that $\ham_{spin}|n\rangle = E_n|n\rangle$. 

\begin{figure}
    \centering
    \includegraphics[width=\textwidth]{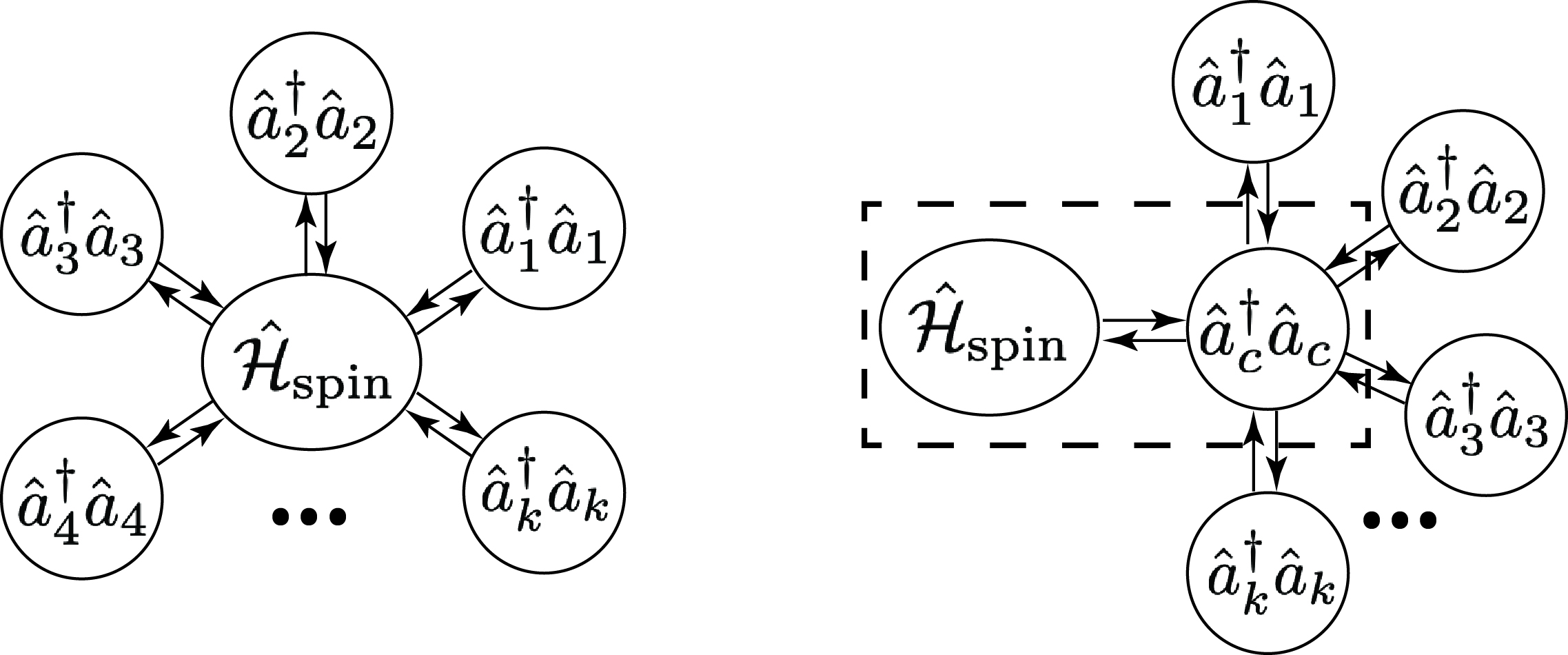}
    \caption{\label{fig:RC} Schematic illustration of reaction coordinate Hamiltonian. Left panel: the spin system directly interacts with all the photon modes in the reservoir. Right panel: the spin system only directly interact with the cavity mode who interaction with other modes in the reservoir, forming an extended system highlighted by the dashed box.}
\end{figure}

In free space, the spin ensemble would interact with all electromagnetic waves. However, when confined within a cavity, the cavity resonator serves as a filter and shields the spin ensemble from non-resonant modes in the environment (reservoir), which reshapes the spectral weight of the cavity EM field according to the physical constraint to center around the resonance of the cavity resonator. This eventually leads to a much stronger coupling between the spin ensemble and the cavity mode in contrast to its interaction with other EM modes. Therefore, it is a reasonable approximation to discard the direct interaction term ($\ham^{int}_{res-spin}$) between the spin ensemble and the reservoir. This redefines the system boundary as schematically illustrated in figure \ref{fig:RC}, in alignment with the reaction coordinate method described in Ref. \cite{Sztrikacs2021,Strasberg2016}. To retain generality, we adopt the same convention in Ref. \cite{Garrison2008}, and impose quantization rules for cavities of arbitrary shape, in which the field is linearly polarized with an inhomogeneous distribution. It is natural to assume that a cavity supports discrete modes while the reservoir, being non-resonant, has a continuous spectrum in the frequency space. Therefore we use subscripts "$c$" on the creation and annihilation operators ($\hat{a}_c^{(\dg)} = \hat{a}^{(\dg)}(\w_c, t)$) to denote all permitted discrete cavity field modes and to differentiate from reservoir operators $\hat{b}^{(\dg)}(\w, t)$ that are continuous function of modes in free space. 

All the measurements presented in this report were conducted within frequency windows smaller than 300 MHz, which is significantly less than the magnitude of $\w_c > 3$ GHz itself. Consequently, it is appropriate to apply the rotating wave approximation (RWA) to the cavity-reservoir interaction and discard terms involving $\hat{a}^\dagger\hat{b}^\dagger$ and $\hat{a}\hat{b}$. This approximation further simplifies the reservoir-cavity interaction Hamiltonian to:
\begin{equation}
\label{Eq:magpol}
\ham^{int}_{cav-res} =  \int\limits_{\w}d\w\; iK(\w)[g^*(\w_c)g(\w)a_c^{\dg}b(\w,t) - g(\w_c)g^*(\w)b^{\dg}(\w,t)a_c]
\end{equation}
where $g(\w_c) = i\sqrt{\mu_0\hbar\w_c/2}$, with $\mu_0$ being the vacuum permeability and $\hbar$ denoting the reduced Planck constant. $K(\w)$ characterizes the coupling between the cavity and reservoir fields and is largely determined by the physical attributes of the cavity.

To avoid over-simplification, we treat the interaction term between the cavity field and the spin ensemble without resorting to either the macro-spin description \cite{Soykal2010} or the Holstein-Primakoff transformation \cite{Zhang2014} when describing the spin ensemble. Taking into consideration of that all spins are localized in real space, we can transform the real-space integration into a summation over all lattice sites. Then we normalize the gauge field following the methodology outlined in Ref. \cite{Flower2019}. After a few straightforward steps detailed in Appendix \ref{appx_cmp}), the interaction term between the cavity field and the spin ensemble is given by:
\begin{equation}
\label{Eq:CS_Ham}
\ham_{cav-spin}^{int} = \sum\limits_{c}g(\w_c)\eta\sqrt{\rho_s}\big(\hat{a}^{\dg}_c\hat{\mu}^{\alpha}_{k_c}+\hat{a}_c\hat{\mu}^{\alpha}_{-k_c}\big)
\end{equation}
where $\rho_s$ stands for the volume density of spins in the sample, and $\hat{\mu}^{\alpha}_{k_c}$ is the magnetic moment in reciprocal space, $k_c$ represents the cavity photon mode. It is worth pointing out that the momentum is conserved, and therefore even in the long wavelength limit, $\hat{\mu}_{0}^{\alpha}$ and $\hat{\mu}_{-0}^{\alpha}$ are differentiate. By using the Hubbard operator for a multi-level system \cite{Garrison2008}: $\hat{\sigma}_{mn} = |m\rangle\langle n|$, where $|m\rangle$ and $|n\rangle$ are pure states, we are then able to express the spin operator with a general momentum ($q$) dependence as: $\hat{\mu}_q^{\alpha} = \sum_{m,n}\langle m|\hat{\mu}_q^{\alpha}|n\rangle\hat{\sigma}_{mn}$ with the summation running over the entire Hilbert space. It is evident that $(\hat{\mu}_q^{\alpha})^{\dg} = \hat{\mu}_{-q}^{\alpha}$ since $\sum_{m,n}\langle m|\hat{\mu}_q^{\alpha}|n\rangle\hat{\sigma}_{mn} = \sum_{m,n}\langle n|\hat{\mu}_{-q}^{\alpha}|m\rangle^*\hat{\sigma}_{nm}^{\dg}$. Thus, it follows that the interaction Hamiltonian in Eq. \ref{Eq:CS_Ham} is hermitian.

Furthermore, in Eq. \ref{Eq:CS_Ham}, we have defined $\eta^2_c = |\int_{V_s}\hat{\textit{e}}^{\alpha}\cdot\vec{B}'(r)e^{i\mathbf{k}_c\cdot \mathbf{r}}dr|^2/[V_s\int_{V_c}|\vec{B}'(r)|^2dr]$ as the filling factor to capture the degree to which the mode volume of the cavity field overlaps with that of the sample\cite{Flower2019}. $\vec{B}'(r)$ is the magnetic component of the EM field with a dimension defined in the supplementary material. $\hat{\textit{e}}^{\alpha}$ is the unit vector along the magnetic spin moment, and the dot product with it simply ensures that only field along the spin component is taken into consideration. For the present experiment, noting that the frequency range ($<$ 4.8 GHz) corresponds to wavelengths ($\lambda > 70$ mm) that are much longer than both the lateral dimensions of our sample ($\sim$ 3 mm) and the inter-spin distance ($\sim$ 10 \AA), we are therefore well within the $k\sim0$ regime, where the exponential phase factor in the filling factor vanishes (in the simpler case of a rectangle box cavity, the filling factor further reduces to a simple ratio between the sample volume, $V_s$, and cavity volume, $V_c$.)

Thus, the total Hamiltonian can then be written explicitly as:
\begin{multline}
\ham  = \sum\limits_c\hbar\w_c\hat{a}_c^{\dg}\hat{a}_c + \int_{\w}\hbar\w \hat{b}^{\dg}(\w,t)\hat{b}(\w,t) + \ham_{spin}\\
 + \int\limits_{\w}d\w\; iK(\w)[g^*_cga_c^{\dg}b(\w,t) - g_cg^*b^{\dg}(\w,t)a_c]\\ + \sum\limits_{c}g_c\eta\sqrt{\rho_s}\big(\hat{a}^{\dg}_c\hat{\mu}^{\alpha}_{k_c}+\hat{a}_c\hat{\mu}^{\alpha}_{-k_c}\big)\\
\end{multline}
where we have used the shorthand notation $g_c = g(\w_c)$ and $g = g(\w)$.

The experimental setup discussed in the present report can be treated as a single sided lossy cavity with an input field ($\hat{b}^\dg_{i}\hat{b}_{i}$) and an output field ($\hat{b}^\dg_{o}\hat{b}_{o}$). All data presented in this study were acquired in the form of scattering parameter $|S_{11}| = |\hat{b}_{o}/\hat{b}_{i}|$. Starting from the Heisenberg equations of motion: $\partial_t\hat{b} = (i/\hbar)[\hat{\mathcal{H}},\hat{b}]$ and $\partial_t\hat{a} = (i/\hbar)[\hat{\mathcal{H}},\hat{a}]$, and following the same steps laid out by ref. \cite{Gardiner1985,Scully1997,Meystre2007,Walls2008}, we obtain a dynamical equation for the cavity field operator:
\begin{equation}
\label{Eq:EqoMa}
    \partial_t\hat{a}_c = -i\w_c\hat{a}_c + \frac{1}{\hbar}\int K^*(\w)\hat{b}_0(\w)e^{-i\w t}d\w + \sum\limits_c\bigg[\frac{K(\w_c)}{\hbar}\bigg]^2\hat{a}_c + \eta\sum\limits_c\sqrt{\frac{\mu_0\omega_c\rho_s}{2\hbar}}\hat{\mu}^{\alpha}_{k_c}
\end{equation}

We proceed by defining: $\hat{b}_i=-\int\hat{b}_0(\w)e^{-i\w(t)}d\w$ $(t<0)$, $\hat{b}_o=\int\hat{b}_0(\w)e^{-i\w(t)}d\w$  $(t>0)$, and adopting the first Markovian approximation to assume that the coupling strength between the cavity field and the environment is frequency independent, in other words $K(\w) = \hbar\kappa_e$. Subsequently, we arrive at:
\begin{align}
    \partial_t\hat{a}_c &= \sum\limits_c\left(\Gamma(\w_c)\hat{a}_c+i\sqrt{\frac{\mu_0\omega_c\rho_s\eta}{2\hbar}}\hat{\mu}^{\alpha}_{k_c}\right)+\sqrt{\kappa_e}\hat{b}_{i} \label{Eq:infield}\\
    \partial_t\hat{a}_c &= \sum\limits_c\left(\Gamma(\w_c)\hat{a}_c+i\sqrt{\frac{\mu_0\omega_c\rho_s\eta}{2\hbar}}\hat{\mu}^{\alpha}_{k_c}\right)-\sqrt{\kappa_e}\hat{b}_{o} \label{Eq:outfield}
\end{align}
where $\Gamma(\w_c) = -i\w_c+(\kappa_i-\kappa_e)/2$. Furthermore, we have introduced an additional Langevin force, characterized by $\kappa_i$, originated from the internal dissipation of the cavity field. This is equivalent of adding an imaginary term to the Hamiltonian that breaks the hermicity of the total Hamiltonian and turns it into an open quantum system coupled to a large reservoir of unknown specifics. More detailed discussions about the reasoning behind this term can be found in \cite{Sachdev1984,Lax1965}. In the present case, this additional term enables us to capture the impedance matching condition between the load (cavity) and drive (generator + circuit), which is never truly perfect in reality. The next step is to find an expression for $\hat{\mu}_k$.To this end, we again resort to the Heisenberg equation of motion:
\begin{equation}
\label{Eq:pdm}
    \begin{aligned}
    \partial_t\hat{\mu}^{\alpha}_{k_c} &= -\frac{i}{\hbar}[\hat{\mathcal{H}}_{spin},\hat{\mu}^{\alpha}_{k_c}] - \frac{i}{\hbar}[\hat{\mathcal{H}}^{int}_{cav-spin},\hat{\mu}^{\alpha}_{k_c}]\\
    &= -\frac{i}{\hbar}[\hat{\mathcal{H}}_{spin},\hat{\mu}^{\alpha}_{k_c}]\\
    & - i\eta\sqrt{\frac{\mu_0\omega_c\rho_s}{2}}\sum\limits_{\substack{m,n\\p,q}}\langle m|\hat{\mu}^{\alpha}_{k_c}|n\rangle\langle p|\hat{\mu}^{\alpha}_{k_c}|q\rangle[\hat{\sigma}_{mn},\hat{\sigma}_{pq}]\hat{a}_c\\
    \end{aligned}
\end{equation}
In order to compute $[\hat{\mathcal{H}}_{spin},\hat{\mu}^{\alpha}_{k_c}]$, we recall the definition $\hat{\mu}_{k_c} = \sum_{m,n}\langle m|\hat{\mu}_{k_c}^{\alpha}|n\rangle\hat{\sigma}_{mn}$ and the fact that our only constraint on $\hat{\mathcal{H}}_{spin}$ is it being hermitian and hence could be diagonalized unitarily. Therefore, we can write the spin Hamiltonian in its diagonal form using the Hubbard operator: $\hat{\mathcal{H}}_{spin} = \sum_{q}\epsilon_q\hat{\sigma}_{qq}$. Consequently, $[\hat{\mathcal{H}}_{spin},\hat{\mu}^{\alpha}_{k_c}] = \sum_{m,n}\hbar\Omega_{mn}\hat{\sigma}_{mn}$, where we have defined $\Omega_{mn} = (\epsilon_m - \epsilon_n)/\hbar$. After performing a Fourier transform on Eq.\ref{Eq:pdm} and collect all the terms on the right hand side gives:
\begin{equation}
    \hat{\mu}^{\alpha}_{k_c} = -i\eta\sqrt{\frac{\mu_0\omega_c\rho_s}{2\hbar}}\sum\limits_{\substack{m,n\\p,q}}\frac{\langle m|\hat{\mu}_{k_c}^{\alpha}|n\rangle\langle n|\hat{\mu}_{-k_c}^{\alpha}|m\rangle[\hat{\sigma}_{mn},\hat{\sigma}_{pq}]}{\w-\Omega_{mn}+i\gamma_{mn}}\hat{a}_c
\end{equation}
where the summation over $m$, $n$, $p$, and $q$ accounts for contributions from all possible excitation/decay process. $\gamma_{mn}$ is an infinitesimal number that has the physical significance of being the spectral line width. The commutation relation in the nominator can be further expanded explicitly as \cite{Garrison2008}: $[\hat{\sigma}_{mn},\hat{\sigma}_{pq}] = \delta_{np}\hat{\sigma}_{mq} - \delta_{mq}\hat{\sigma}_{pn}$. To proceed, consider an arbitrary eigen-state of the spin system $|\psi\rangle$. By definition we have $\hat{\sigma}_{mm}|\psi\rangle = \langle m|\psi\rangle |m\rangle$, where $\langle m|\psi\rangle$ is the expansion coefficient. Within the mean field framework, where no mode-mode interaction exists, the spectral weights in equilibrium are entirely governed by the Boltzmann distribution, therefore $\langle m|\psi\rangle = e^{-\beta E_m}/Z$ and $\hat{\sigma}_{mm} = N_m$. The last term of Eq.\ref{Eq:EqoMa} hence gives:
\begin{equation}
\label{eqn:Gw}
\begin{split}
&i\eta^2\frac{\mu_0\omega_c\rho_s}{2}\sum\limits_{m,n}\frac{\langle m|\hat{\mu}_{k_c}^{\alpha}|n\rangle\langle n|\hat{\mu}_{-k_c}^{\alpha}|m\rangle}{\hbar(\omega-\Omega_{mn})+i\hbar\gamma_{mn}}(N_m-N_n)\\
&= i|G(\w_c)|^2\chi_{k_c}^{\alpha\alpha}(\w)
\end{split}
\end{equation}
where we have defined $G(\omega_c) = \eta\sqrt{\mu_0\omega_c\rho_s/2}$ and invoked Eq.\ref{Eq:gsus} with the definition $\mathcal{\hat{A}} = \mathcal{\hat{B}} = \hat{\mu}_k^\alpha$. Finally, by using the Fourier transform Eq.\ref{Eq:infield} and Eq.\ref{Eq:outfield}, we arrive at a simple expression for $|S_{11}|$ after some straightforward algebra:
\begin{equation}
\label{Eq:s11x}
|S_{11}| = \left| 1 + \frac{2\kappa_e}{i(\omega-\omega_c)-(\kappa_e+\kappa_i)+i|G(\w_c)|^2\chi_{k_c}^{\alpha\alpha}(\w) }\right|
\end{equation}
which includes contributions from all possible transitions of a diagonal susceptibility of the spin system. Given that no specifics constraints have been imposed on the spin ensemble, Eq.\ref{Eq:s11x} should, in principle, apply to all magnetic materials in the linear response regime. Rewriting the generalized susceptibility in explicit real and imaginary parts ($\chi_{k_c}^{\alpha\alpha}(\w) = \textrm{Re}\big[\chi_{k_c}^{\alpha\alpha}(\w)\big] + i\cdot\textrm{Im}\big[\chi_{k_c}^{\alpha\alpha}(\w)\big]$) subsequently yields:
\begin{equation}
\label{Eq:s11xrxi}
|S_{11}| = \left| 1 + \frac{2\kappa_e}{i(\omega-\omega_c+|G(\w_c)|^2\textrm{Re}\big[\chi_{k_c}^{\alpha\alpha}(\w)\big])-(\kappa_e+\kappa_i+|G(\w_c)|^2\textrm{Im}\big[\chi_{k_c}^{\alpha\alpha}(\w)\big]) }\right|
\end{equation}
it is evident from this expression that the real part of the susceptibility is responsible for shifting the resonant frequencies of the cavity-spin system, whereas the imaginary part of it influences the damping rate and hence the overall line width.

\begin{figure*}
    \centering
    \includegraphics[width=\textwidth]{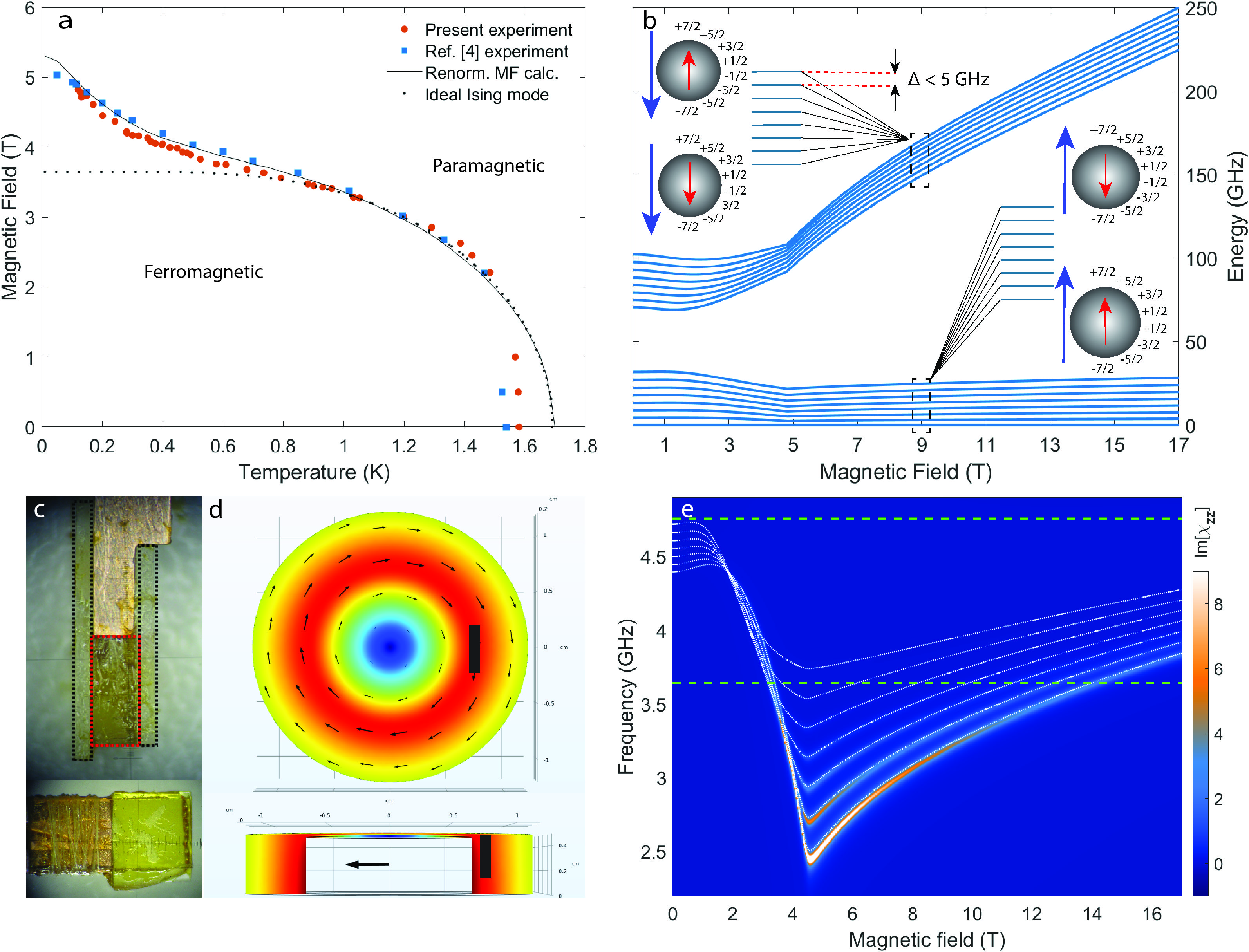}
    \caption{\label{fig:setup} (a) Magnetic phase diagram of LiHoF$_4$. The red circles and blue squares are experimental data of the magnetic phase boundaries from, respectively, this work and Ref. \cite{Bitko1996}. The black doted and solid curves are MFT values of, respectively, an ideal Ising model and LiHoF$_4$ with renormalized spins\cite{Ronnow2007}. (b) The hyperfine splitting of the first two electronic spin states of LiHoF$_4$ by MFT calculation. The insets illustrates the product-states of the electro-nuclear spins.  (c) Photos of the LiHoF$_4$ sample (inside the red dashed box) in crystallographic a-c plane (upper panel) and a-b plane (lower panel) mounted between two sapphire substrates (encompassed by black dashed lines). (d) Top view (upper panel) and side view (lower panel) of the intensity (color map) and the magnetic polarization (small black arrows) of the cavity photon field from FEM calculation using COMSOL$^\text{\textregistered}$. The large black arrow in the lower panel indicates the crystallographic c-axis of the mounted sample. The black rectangle in both panels indicates the sample location to scale.(e) Color map: the imaginary part of calculated longitudinal magnetic susceptibility of LiHoF$_4$ at 100 mK. The white dashed lines plotted on top are the first seven electro-nuclear spin transitions from MFT calculation. The two green dashed lines indicates the bare resonant frequencies of the microwave cavity used in the experiment.}
\end{figure*}

\subsection*{Magnetic model of LiHoF$_4$}
The total Hamiltonian of LiHoF$_4$ then reads:
\begin{equation} 
\label{Eq:fHam}
\ham = \sum_i[\ham_{CF} + g_L\mu_B\hat{J}_i\cdot\vec{B}_\perp + g_N\mu_N\hat{I}_i\cdot\vec{B}_\perp + A\hat{I}_i\cdot\hat{J}_i] + \sum_{i>j} [\mathcal{J}_{ex} \hat{J}_i\cdot\hat{J}_j + (g_L\mu_B)^2\hat{J}_{i}\overline{\overline{\mathcal{D}}}\hat{J}_j]
\end{equation}
where $g_{L/N}$ are the electronic/nuclear Land\'e factor, $\mu_{B/N}$ are Bohr/nuclear magneton. $\hat{J}_i$ is the electronic spin on lattice site $i$. $\overline{\overline{\mathcal{D}}}$ is the dipole interaction tensor. The summation in Eq. \ref{Eq:fHam} are carried over all sites ($i$, $j$). 

Finding the exact eigen-states to a many-body system like LiHoF$_4$ is in general difficult, and a commonly used approximation to simplify the problem is the mean field approximation (MFA), which is introduced through $\hat{J}_i = \langle J\rangle + \delta \hat{J}_i$. Consequently, the aforementioned many-body Hamiltonian is reduced to a single-ion Hamiltonian by discarding $\delta \hat{J}_i \cdot \delta \hat{J}_j$ and $\delta \hat{J}_i\overline{\overline{\mathcal{D}}}\delta \hat{J}_j$ in the spin-spin interaction terms\cite{Kovacevic2016}. The resulting single-ion Hamiltonian is then numerically diagonalized in a self-consistent manner, yielding the mean field (MF) eigen-states and eigen-energies. The ground state doublet of the electronic spin states in this solution are now split by the hyperfine interaction and is shown in figure \ref{fig:setup}(b): The two "bundles" of hyperfine states, represented by the solid blue lines, each correspond to one of the electronic ground state doublet. The cartoon insets depict the relative orientations between the electronic and nuclear spin moments.

The single-ion dynamical magnetic susceptibility is calculated using Eq.\ref{Eq:gsus} with the MF eigen-states obtained in the previous step. Despite the inclusion of nuclear Zeeman interactions, the largest amplitude of AC magnetic susceptibility remains aligned with the easy axis (c-axis) of LiHoF$_4$, which echos the previous finding in Ref. \cite{McKenzie2018}.

\subsection*{\label{sec:exp} Experimental results}
\begin{figure*}
\centering
    \includegraphics[width=\textwidth]{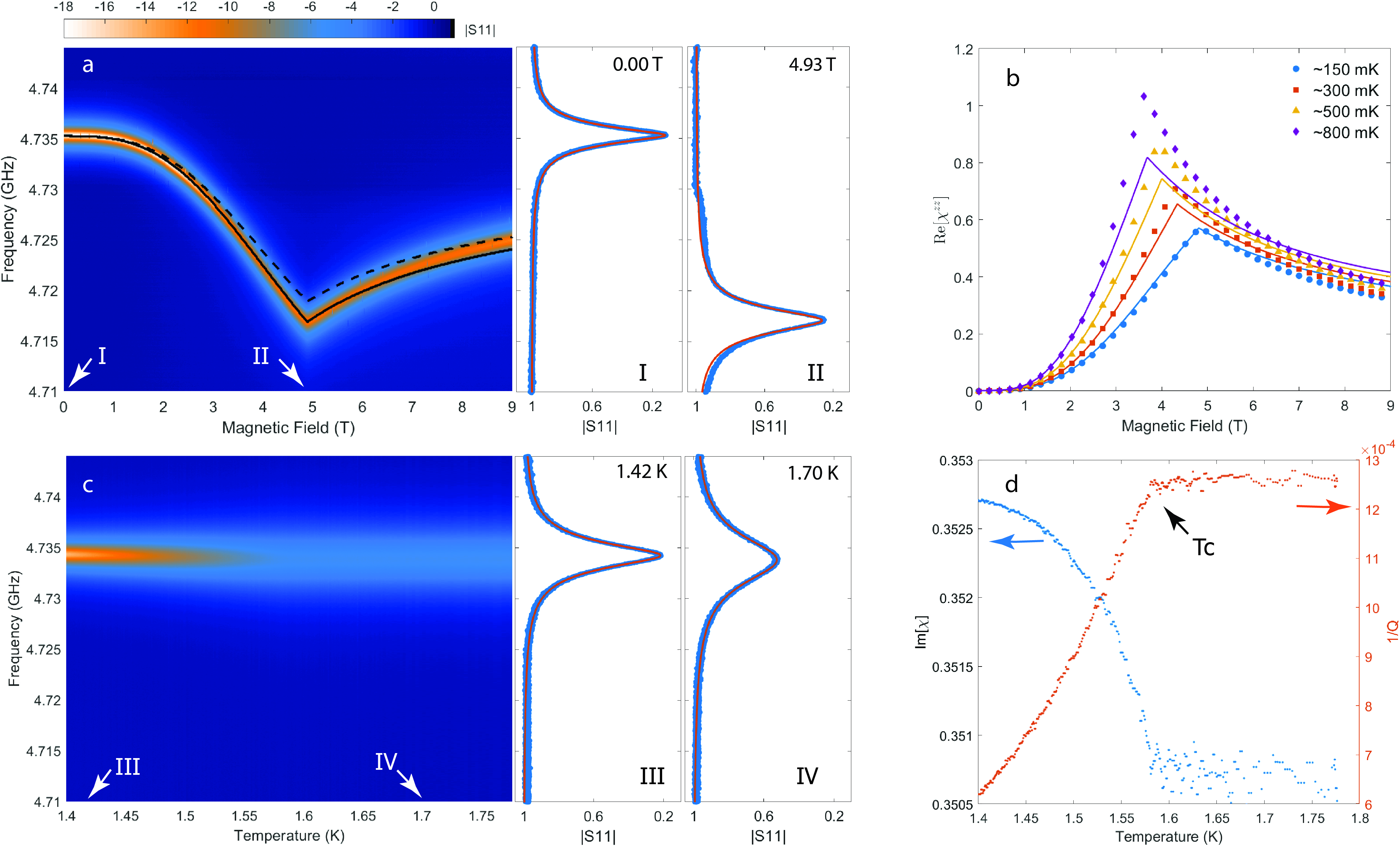}
\caption{\label{fig:data_off}(a) Left panel: measured $|S_{11}|$ in dB at 150 mK as a function of external magnetic field and frequency. Dashed line: theoretical values calculated using FEM, MFT and input-output formalism. Solid line: theoretical values with calculated filling factor scaled by 1.06. Middle and right panel: individual fits to measured $|S_{11}|$ at 0.00 T and 4.93 T (location indicated by arrows in the left panel) (b) Fitted values (markers) from off-resonance measurements and calculated values (lines) of the real part of the longitudinal susceptibilities as a function of magnetic field at $\sim 150$ mK (blue), $\sim 300$ mK (red), $\sim 500$ mK (yellow), and $\sim 800$ mK (purple). (c) Left panel: measured $|S_{11}|$ at 0.00 T as a function of temperature and frequency. Middle and right panel: individual fits to the measured $|S_{11}|$ at 1.42 K and 1.70 K (location indicated by arrows in the left panel). (d) The imaginary part of the susceptibility (blue dots) and the Inverse of the quality factor (red dots) obtained yb fitting to the data shown in (c).}
\end{figure*}

We first focus on the off-resonance data, where the cavity mode is sufficiently detuned from the electro-nuclear spin transitions, direct hybridization between the cavity and electro-nuclear spin transitions are negligible. We present measured $|S_{11}|$ from this regime at 150 mK as a function of frequency and external magnetic field in Figure \ref{fig:data_off}(a), obtained with a cavity frequency of 4.74 GHz (above the nuclear spin transitions in LiHoF$_4$ across the entire magnetic field range). As expected, no evidence of direct hybridization, such as branching, is seen in the result. In the meantime, apart from minor changes to the cavity line width, the Coupling between the cavity and the sample does result in a dip with a kink at the critical field of the quantum phase transition ($\sim 5$ T) in the cavity mode as a function of the magnetic field. To compare with our model, we reproduce the experimental data by using Eq.\ref{Eq:s11x} with the MF eigen-states of Eq.\ref{Eq:fHam} applied to Eq.\ref{Eq:gsus}. We plot the theoretical results in dashed black line, overlaid on top of the experimental data in figure \ref{fig:data_off}(a). The shift in resonance frequency is proportional to the squared coupling constant $G(w)$, which in turn is proportional to the filling factor ($\eta$) we calculated from FEM analysis via COMSOL. Remarkably, without adjusting any parameters neither in the MF-RPA model nor in the cavity coupling constant, the prediction (dashed black line) comes very close to the measurement. Scaling the coupling constant by a factor 1.06 makes the agreement perfect (black solid line) up to the critical field. Our simulation captures well the essential features of the experimental data and hence the kink in the resonance frequency is understood as a manifestation of the strongly elevated magnetic susceptibility at the QCP.

\begin{figure}
    \centering
    \includegraphics[width=0.43\textwidth]{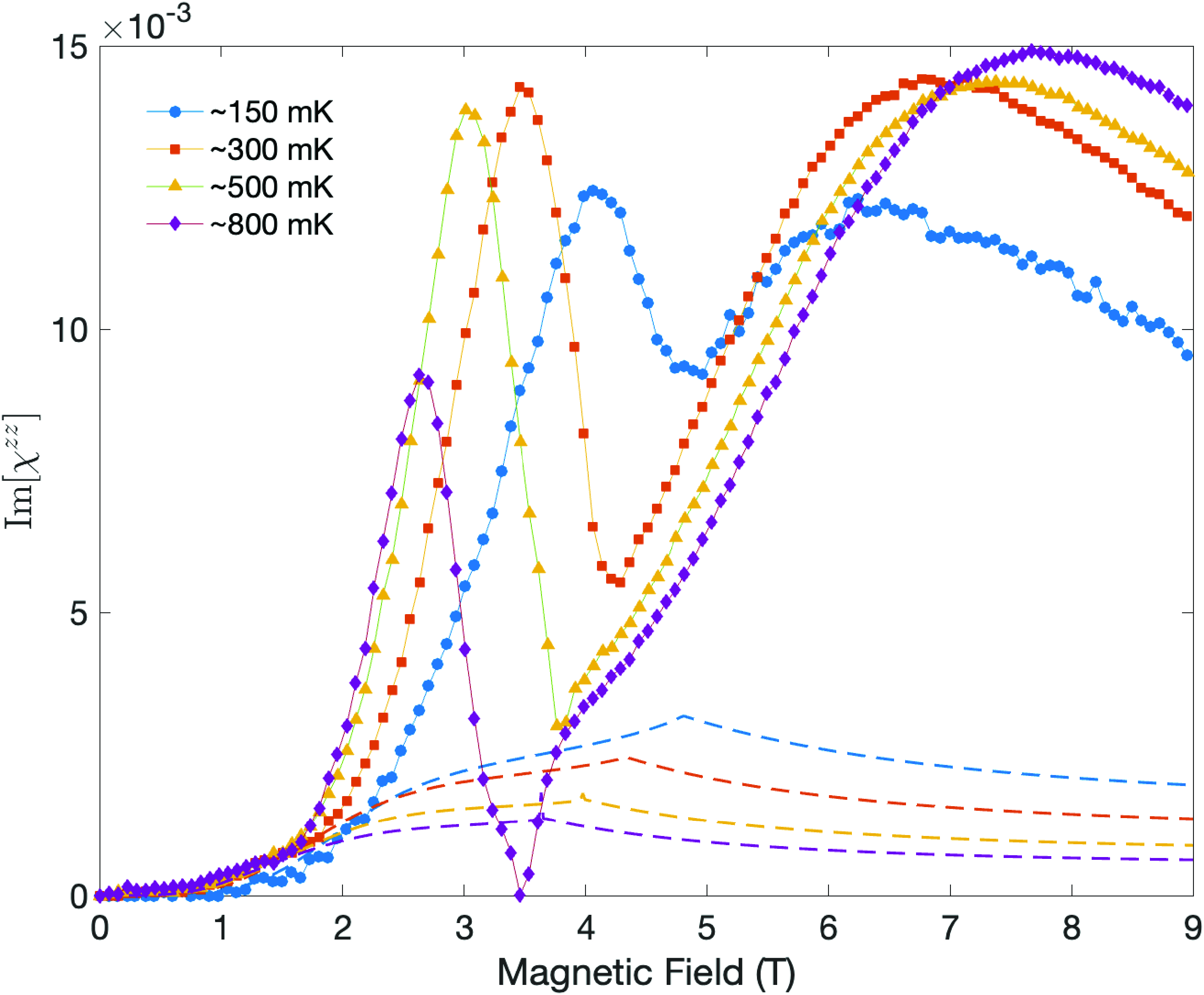}
        \caption{\label{fig:data_off_imag}Fitted values (markers) from off-resonance measurements and calculated values (lines) of the imaginary part of the longitudinal susceptibilities as a function of magnetic field at $\sim 150$ mK (blue), $\sim 300$ mK (red), $\sim 500$ mK (yellow), and $\sim 800$ mK (purple). }
\end{figure}

In addition to the absence of direct hybridization between the cavity mode and the electro-nuclear transitions in the off-resonance setup, both the real and imaginary parts of the magnetic susceptibility of the spin ensemble are very flat, as functions of frequency, in the vicinity of the cavity mode. Therefore, it is a good approximation to consider $\chi_{k_c}^{zz}(\w)$ frequency independent within a narrow frequency window, akin to conventional AC susceptibility measurements. This allows us to fit the measured $|S_{11}|$ in off-resonance experiments using Eq.\ref{Eq:s11xrxi}. Since the hamiltonian commutes, at zero external magnetic field, with the spin projector along the Ising axis for an Ising system, its logitudinal magnetic susceptibility vanishes according to Eq. \ref{Eq:gsus}. Therefore, in fitting the data, we first establish the baseline by fitting $|S_{11}|$ at zero magnetic field using Eq. \ref{Eq:s11x} with $\chi^{zz}_{k_c}(\w)$ set to zero. This yields the external ($\kappa_e$) and internal ($\kappa_i$) dissipation rates of the cavity \cite{Schuster2010,Abe2011,McKenzie2019}, which are $\kappa_e/2\pi\approx556$ kHz and $\kappa_i/2\pi\approx381$ kHz respectively. Subsequently, the remaining measurement of $|S_{11}|$ at finite fields are fitted while keeping $\kappa_e$ and $\kappa_i$ fixed at these baseline values. Two representative fits at zero field and the critical field are displayed in the right panels of figure \ref{fig:data_off}(a), where the blue dots represent the experimental data and orange lines are the corresponding fits. The extracted susceptibilities as functions of magnetic field at four selected temperatures (150 mK, 300 mK, 500 mK, and 800 mK) are given in figure \ref{fig:data_off}(b) with color coded markers. The evolution of the extracted susceptibility as a function of magnetic field and temperature is qualitatively reproduced by the MFT model (solid lines). The experimental susceptibility grows more than the prediction at elevated temperatures, which is likely linked to the negligence of fluctuations in the MF approximation. 

\begin{figure*}
\centering
\includegraphics[width=\textwidth]{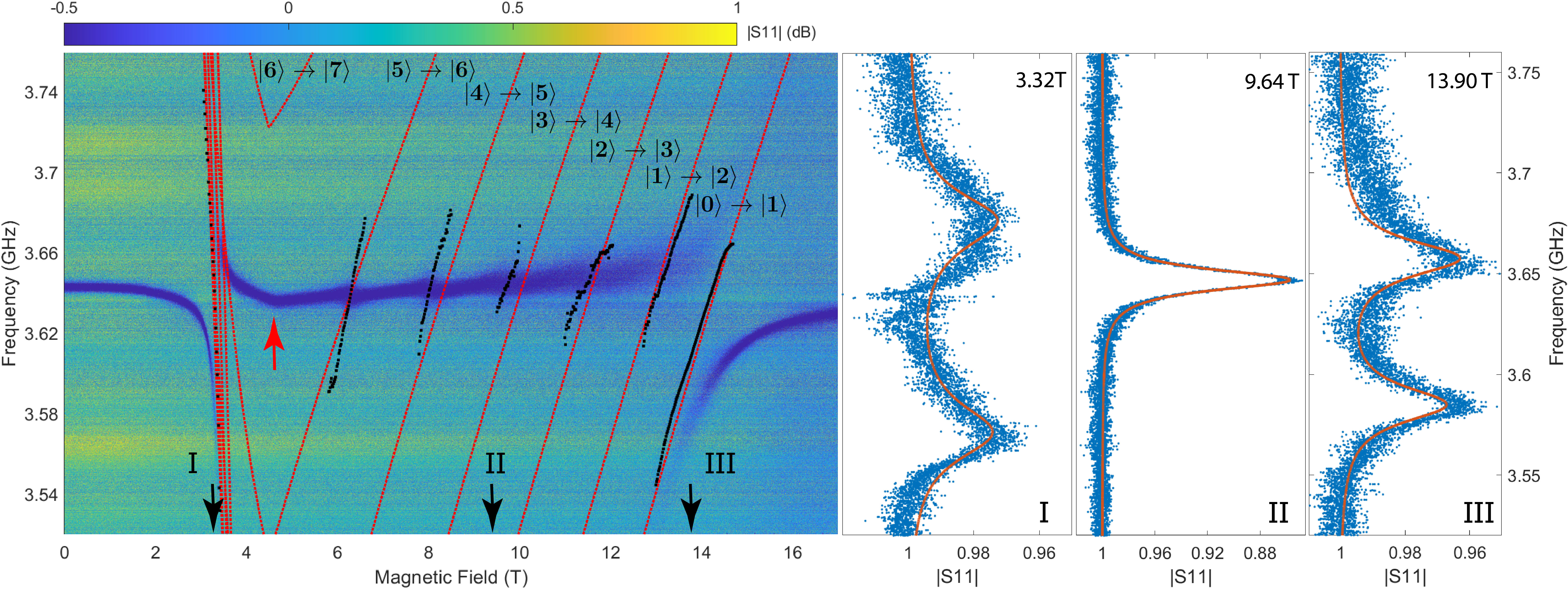}
\caption{\label{fig:data_on}Measured $|S_{11}|$ in dB as a function of magnetic field at 130 mK. Red dashed lines: calculated excitation spectra by the MFT. Black dots: fitted excitation spectra. The red arrow indicates the cusp in the trace of cavity resonance frequency that was used to infer sample temperature. The right three panels show individual fits to the experimental data (blue dots) at linear scales and various field locations indicated by the three black arrows on the left panel.}
\end{figure*}

As the phase transition between ordered and polarized paramagnetic states is clearly visible as a kink in the resonance frequency, we are able to obtain the field-temperature phase diagram by repeat the off-resonance measurement at various temperatures. While this method is very effective at low temperatures, it becomes unreliable at high temperatures because: first, the phase boundary becomes nearly parallel to the field axis, causing the measurement trajectory to intersect the phase boundary at increasingly shallow angles; Secondly, strong thermal fluctuations in the spins lead to significant damping, resulting in substantially broadened cavity mode. Consequently, near the zero-field-critical-temperature, we perform temperature-sweeping measurements at fixed magnetic fields to track the phase boundary. In figure \ref{fig:data_off}(c), we show measured $|S_{11}|$ at zero field in the left panel, while two individual sets of data with fits taken at temperatures on either side of the critical temperature are given in the right panel. Unlike field-sweeping measurement, data from temperature-sweeping measurement does not display any significant frequency shift but rather a pronounced line-width broadening. In figure \ref{fig:data_off}(d) we show both the fitted imaginary part of AC susceptibility as well as the inverse of the loaded quality factor, defined as $1/Q = \text{FWHM}/f_0$, where FWHM stands for full-width-half-max and $f_0$ is the peak frequency. A clear inflection point occur at the phase transition can be seen in both sets of curves. Combining the experimental data from both field- and temperature-sweeping measurements, we are able to map out the entire phase boundary, shown in figure \ref{fig:setup}(a) in solid circles. Our results are in very good agreement with previously published data \cite{Bitko1996}, given in solid squares in the same figure.

In addition to the the real part of the longitudinal magnetic susceptibilities shown in figure \ref{fig:data_off}(b), the imaginary part is also obtained from the fit and presented in figure \ref{fig:data_off_imag} using dashed lines and matching colors. Contrary to the case with the real part, it is evident that the two sets of data of the imaginary part not only disagree with each other quantitatively but also qualitatively, exhibiting different behaviors as functions of the magnetic field. A prominent and consistent feature seen in these experimental data is the double peak surrounding the QCP, which are not present in the calculated results. While the origin of these peaks is unclear to us, the failure of the MFT in capturing them suggest that they are likely related to fluctuations. Furthermore, it is worth noting that they are not expected to be related to the soft mode observed in Ref. \cite{Libersky2021}, which manifested itself as also dual peaks in dissipation surrounding the QCP. Despite the similarity on the surface, our data present in figure \ref{fig:data_off_imag} is taken at a frequency (4.9 GHz), which is not only far above in energy the expected collective soft mode but also the first manifold of the single-ion electro-nuclear spin excitations. Therefore, the cavity mode presented here never intercepts the soft mode, and hence the peaks observed in Im[$\chi^{zz}$] in experiments are not expected to be related to the latter.

Next we focus on the on-resonance measurements where the resonance frequency of the cavity cross through those of the electro-nuclear spin excitations, leading to direct hybridization between the cavity photons and the spin transitions. We present the measured $|S_{11}|$ for a cavity with resonance frequency 3.64 GHz as a function of frequency and field, at 100 mK in figure \ref{fig:data_on}. Several features are visible in the data: (1) similar to the off resonance data, the kink (the red arrow in Fig. \ref{fig:data_on}) at 4.75 T marks the phase transition and can now be used to for direct determination of the sample temperature inside the cavity in presence of the cavity photon field, yielding 130$\pm$5~mK. (2) apparent mode splittings are seen at the two ends of the field axis (3.32 T and 13.90 T). The latter of the two has a very faint upper branch, but nonetheless exhibits a visible avoid-crossing pattern. It is also noticeable that the upper branch at 13.90 T has a broader linewidth than the lower branch, the detailed mechanism is not entirely clear. But one possible explanation is the damping effect from magnon modes in proximity, which is absent for the lower branch as it is below the lowest excitation mode. (3) All other features that appear in between the two remain single-peaked. The occurrence of these features is consistent with the prediction by the CMP theory of perturbation, or bifurcation, of the cavity mode as a result of spin-photon mode hybridization at various strengths\cite{Tavis1967, Tavis1968, Gardiner1985, Schuster2010, Flower2019}. Furthermore, these mode crossings below and above the critical field nearly coincide with those predicted by the MFT (dashed red lines in fig. \ref{fig:data_on}). 

We proceed to fit the on-resonance experimental data, where the magnetic susceptibility can no longer be considered frequency independent due to direct mode hybridization. However, noting that all the mode crossings above the critical field are well separated along the frequency axis, as seen in the left panel of fig. \ref{fig:data_on}. At each of these crossings, only one spin transition directly hybridizes with the cavity mode and exhibits appreciable frequency dependence while all other transitions are sufficiently detuned from the cavity mode. Therefore, by applying the same reasoning as in the previous analysis of off-resonance data, we can treat all other transitions as frequency independent in the close vicinity of each of these crossings. Consequently, we fit the experimental data piece wise at each mode crossing using Eq. \ref{Eq:s11xrxi}, retaining frequency dependence for only the relevant magnon mode. For the avoid-crossing pattern below the critical field, since the MFT predicted that all the electro-nuclear spin transitions at this field nearly coincide with each other (see figure \ref{fig:setup}(e)) and there is only one branching feature visible in the experimental result, we thus fit this part of the data using the same fitting function but with the assumption of all relevant transitions being degenerate. The fitting function used explicitly reads:

\begin{equation}
\label{Eq:s11_fit}
|S_{11}| = \left| 1 + \frac{2\kappa_e}{i(\omega-\omega_c) - (\kappa_e+\kappa_i) + \chi_0' + \frac{\tilde{G}_{mn}^2}{i(\omega-\Omega_{mn})-\gamma_{mn}}} \right|
\end{equation}
where the last term in the denominator describes the magnon mode that directly hybridizes with the cavity mode, and $\chi_0'$ contains contributions from all other spin transitions to the magnetic susceptibility. The three fitting parameters relevant to the on-resonance mode are: 1. $\Omega_{mn}$, the frequency of the electro-nuclear spin transition between $|m\rangle$ and $|n\rangle$ that is on-resonance with the cavity mode frequency; 2. $\tilde{G}_{mn}^2 = |G(\w_c)|^2\langle m|\tilde{\mu}_{k_c}^z|n\rangle\langle n|\tilde{\mu}_{-k_c}^z|m\rangle(N_m-N_n)$ characterizes the coupling strength between the cavity mode and magnon transition; and 3. $\gamma_{mn}$ represent the spectral line width of the corresponding magnon transition. In the right panel of figure \ref{fig:data_on}, we show such examples of individual measurements and fits at three selected magnetic fields that are marked by Roman numerals in the left panel. Panel ''I" and ''II" clearly show branching, or the avoid crossing pattern, that is typical for the strong coupling regime while panel ''II" displays only a single peak, indicating a weak coupling between the cavity field and the spin system.

The electro-nuclear spin transition frequencies extracted from the fits are shown in black points in the left panel of Fig. \ref{fig:data_on}. To compare it to theoretical predictions, we plot the MFT results of the electro-nuclear transitions between single-ion states $|m\rangle$ and $|n\rangle$ ($\Omega_{mn}$) in red dashed lines in the same figure with corresponding labels. The agreement between the calculated and experimental data is found to sensitively depend on the angle of the external magnetic field in the crystallographic ab-plane of LiHoF$_4$ (see appendix \ref{appx_CF} for more details). In order to reach the agreement shown in the figure, an angle of 13$^{\circ}$ away from the crystallographic a-axis within the ab-plane was used for the MFT calculation. With this small angle, a close agreement between the extracted frequencies and the theoretical predictions is observed, both in the locations of the crossings and in the frequency-field slopes.

\begin{figure*}
\centering
    \includegraphics[width=\textwidth]{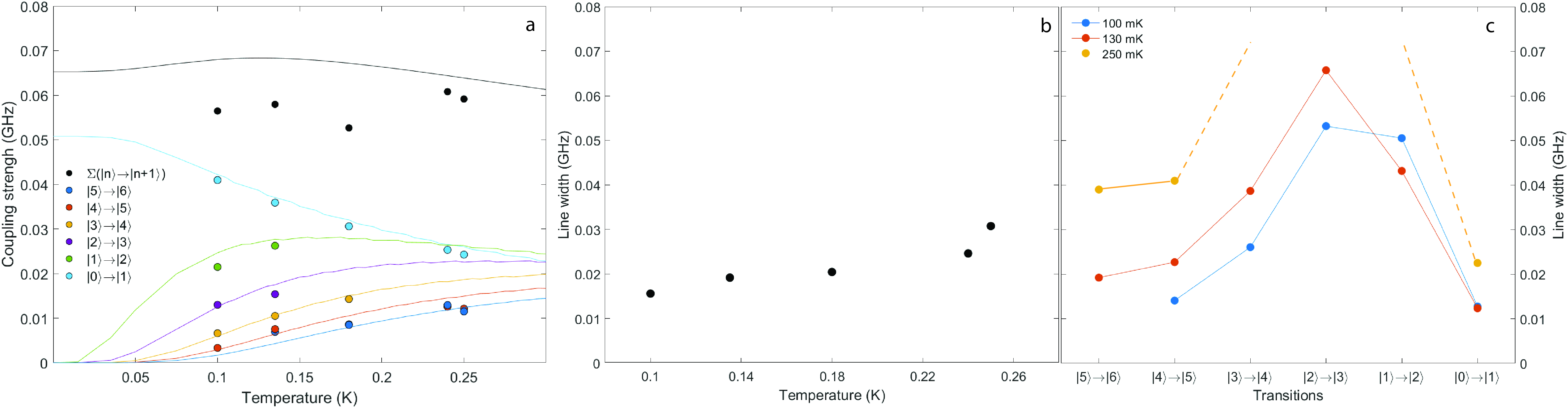}
\caption{\label{fig:fit_param}(a) Fitted individual (colored markers) and collective (solid black)  coupling strengths of electro-nuclear excitation spectra of LiHoF$_4$ at various temperatures. Solid curves are calculated values for individual magnon mode. Different transitions are color coded and matched for both experimental and calculated values. (b) Fitted spin line width from experimental data. (c) Fitted spin line width of the collective mode above critical field.}
\end{figure*}

We also fit the experimental data for coupling strengths at five different temperatures, and the results are presented in figure \ref{fig:fit_param}(a), where the lines represent calculated values and the markers are extracted from the fits to the data. The different magnon transitions are color coded and consistent between the experimental and the theoretical results. For the coupling strength below the critical field, the extracted values from the experimental data are plotted using solid black markers while the MFT calculated values are presented using a solid black line. It is labelled as $\sum(|n\rangle\rightarrow|n+1\rangle)$ since it contains contributions from all the nuclear spin transitions in the first electronic spin manifold. More specifically, it is taken as $\sqrt{\sum_n \tilde{G}_{n,n+1}^2}$. It is clear from the figure that the fitted values appear to vary very little from each other at different temperatures. This behavior is qualitatively consistent with the theoretical prediction (solid black line in Fig. \ref{fig:fit_param}). However, there is a quantitative difference of $\le$15\% between the two as the MFT at all temperatures overestimates the collective coupling strengths. 

In contrast, excellent agreement is reached for the isolated magnon transitions above the critical field at all five temperatures, further confirming our assignment of the magnon transitions to the crossing features in the raw data. It also elucidates the mechanism underlying their varying appearances: When the cavity is driven at low power as is in our case, the interaction between the probe field and the spin ensemble is kept in the perturbative regime. The occupation numbers of spin states therefore do not deviate far from the Boltzmann distribution. 

The MFT calculation shows that the variation in values of the off diagonal matrix elements ($\langle m|\hat{\mu}_{k_c}^{\alpha}|n\rangle$ and $\langle n|\hat{\mu}_{-k_c}^{\alpha}|m\rangle$) in Eqn. \ref{eqn:Gw} is of four orders of magnitude less than the occupation numbers of states throughout the temperature range relevant to the present experiment. Consequently, the temperature dependence of the coupling strength ($G(\omega_c)$) between the photon field and a magnon transition are almost entirely dictated by the occupation number difference between the two involved spin states. This results in the coupling strength changing from weak coupling regime for higher magnon transitions to strong coupling regime for lower magnon transitions, as evidenced by panel ``II" and ``III" of figure \ref{fig:data_on}, where the former magnon transition is between $|3\rangle$ and $|4\rangle$ while the latter is between $|0\rangle$ and $|1\rangle$. Moreover, as the temperature lowers, the higher spin states become depleted, which can weaken the coupling strength significantly enough that the corresponding crossing feature is no longer detectable. This is the case for the experimental data at 100 mK that corresponds to the magnon transition between $|5\rangle$ to $|6\rangle$. Conversely, several fits for the lower transitions at higher temperatures are also absent. In contrary to the case of higher transitions at low temperatures, these lower transitions at high temperatures evade reliable and meaningful fits primarily due to low cooperativity ($C_{mn} = \tilde{G}_{mn}/\sqrt{\gamma_{mn}(\kappa_e + \kappa_i)}$) as a result of thermal broadening.

\begin{figure*}
\centering
\includegraphics[width=\textwidth]{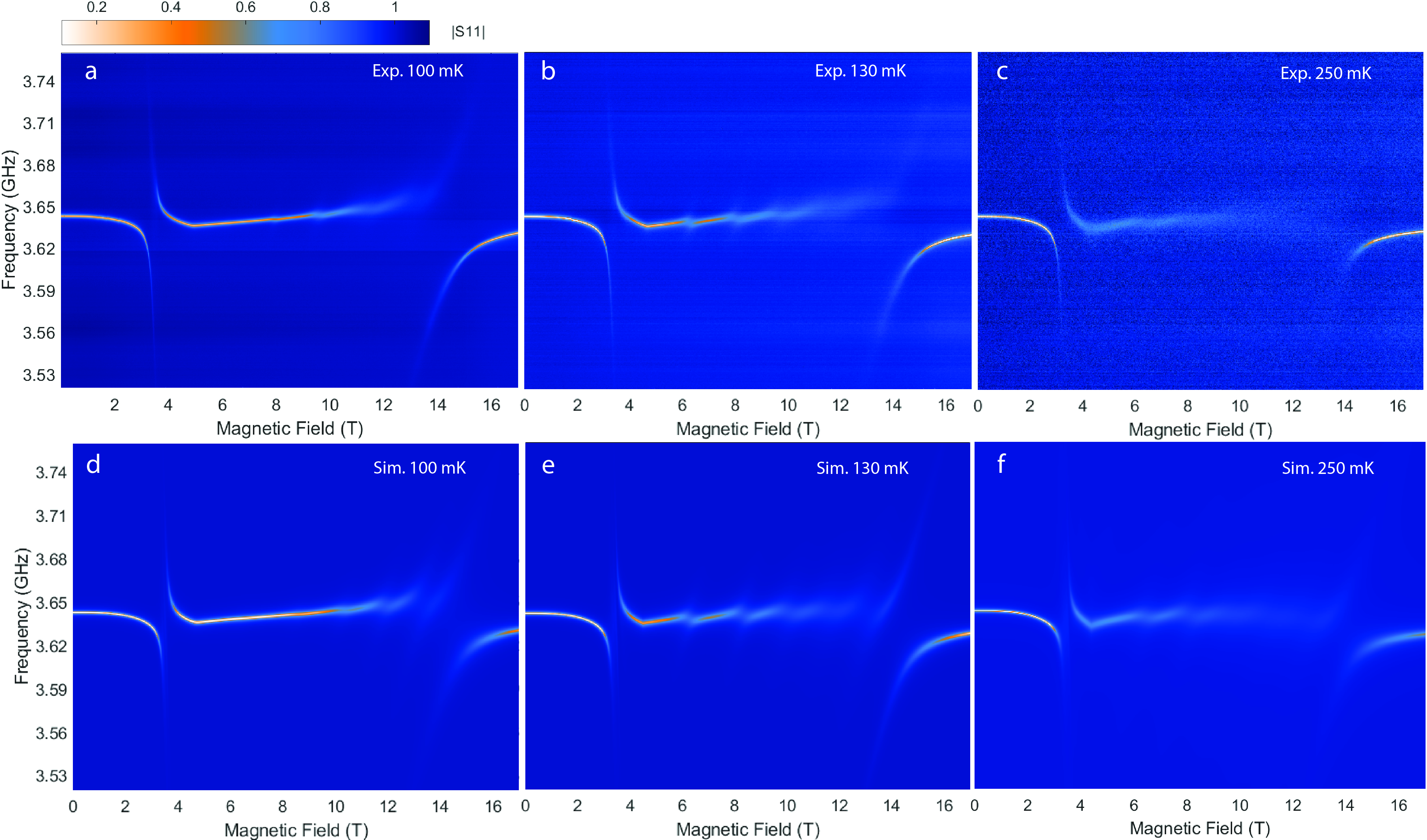}
    \caption{\label{fig:sim-exp} (a-c) Measured $|S_{11}|$ across the quantum critical point at 100 mK (a), 130 mK (b), and 250 mK (d-f) Simulated data of (a-c) in the same order.  For 100~mK we assumed $\gamma_{56}=\gamma_{45}$ and for 250~mK we used $\gamma_{12}=\gamma_{23}=\gamma_{34}=70$~MHz.}
\end{figure*}

The thermal broadening of the spin spectral line width is also directly quantified in figure \ref{fig:fit_param}(c), where we plot the fitted values of the spectral line widths ($\gamma_0/2\pi$) above the critical field at three different temperatures between 100~mK and 250~mK. The solid lines serve as visual guides only while the dashed lines represent our hypothesized trend. For each magnon transition, it clearly shows a trend of broadening toward higher temperatures. In the meantime, the line widths for different magnon transitions at the same temperature show a bell shaped curve where the line widths are the lowest for both the highest and lowest transitions while being elevated for those in between. In consideration of the distance in energy among these magnon transitions, as shown in figure \ref{fig:data_on}(a), the broadening of middle transitions is likely due to the mode-mode coupling where a magnon transition is damped by another nearby transition. With the fitted values of coupling strengths and spin line widths, and previously obtained values of the cavity dissipation rates ($\kappa_e$ and $\kappa_i$), we estimate the cooperativities for the two branching features at 3.32 T and 13.90 T to be, respectively $C\sim 14$ and $C\sim 11$, satisfying the conventional criteria of strong coupling, $C > 1$\cite{Schuster2010}.

\begin{figure}
    \centering
    \includegraphics[width = \textwidth]{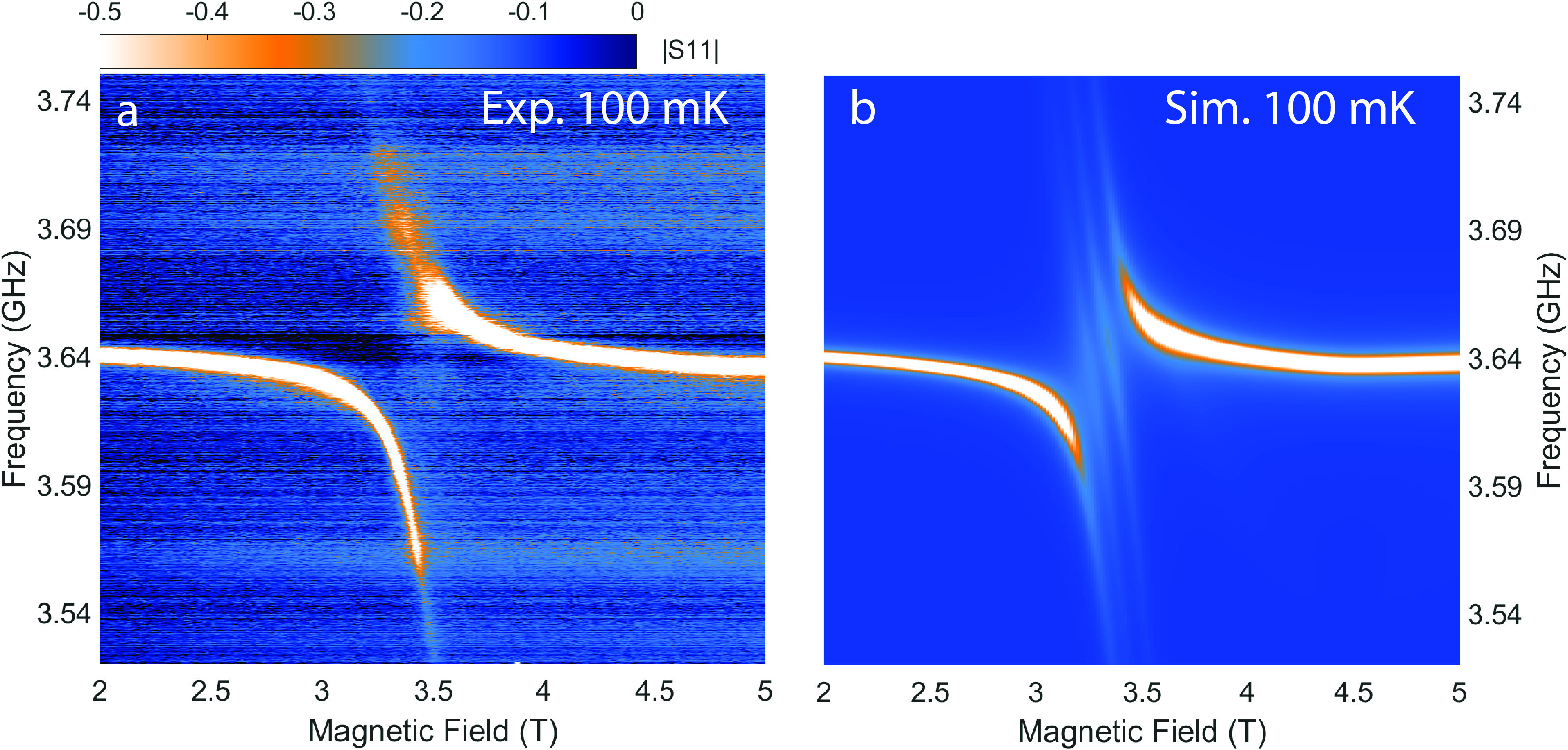}
    \caption{\label{fig:data_on_zoom}(a) Measured $|S_{11}|$ near 3.5 T at 100 mK. (b) Simulated data of (a).}
\end{figure}

Having analyzed individually each mode crossing as function of magnetic field and temperature, we now turn to compare our experimental data to the full MFT simulation obtained from eq. (13), shown in figure \ref{fig:sim-exp} where panel (a), (b), (c) are experimental data and (d), (e), (f) are simulations. For these simulations we used the experimentally extracted line widths (Figure 6(c)) with the exception for the three transitions at 250~mK where reliable fitting was hindered by the broadening. For these transitions, the line widths are set to 70 MHz, which is the largest value obtained from fitting to the experimental data at 130 mK. It is evident from figure \ref{fig:sim-exp} that the simulation adequately reproduces all key features observed in the raw data. These include the kinks in the trace of the cavity mode at the critical field, the increasing coupling strength toward lower magnon transitions, and the enhanced broadening at higher temperatures. The overall agreement is remarkable considering that besides the aforementioned minor corrections to the filling factor and the orientation of the transverse field, no other parameter was adjusted in the model.

However, two minor discrepancies between the calculated and experimental data emerge when zooming in on the branching features of the cavity mode below the critical field, Figure \ref{fig:data_on_zoom}. The first discrepancy is that, while the MFT predicts that the underlying electro-nuclear spin transitions are non-degenerate in the region, as shown in Fig. \ref{fig:data_on}(a), resulting in the compact multi-crossing pattern in Fig. \ref{fig:data_on_zoom}(b), these features are largely subdued, if not completely absent, from the experimental result. Attempts with scaling the transition energies, varying transition line-widths could not resolve this difference. We therefore conjecture that a collective mode, arising from the higher order effects of spin-spin correlation neglected by the MFT, may be at play. This underscores the need for extending beyond the zeroth-order calculation to achieve a comprehensive understanding of LiHoF$_4$ as an interacting spin ensemble. Such an extension is beyond the scope of this report and is left for future theoretical investigations. The second discrepancy is again the comparatively heavier damping of the upper branch in the avoid crossing pattern as seen in fig. \ref{fig:data_on} around 14 T, this is clearly not captured by our simulation by comparing fig. \ref{fig:data_on_zoom}(a) and (b). A possible explanation, similar to the case shown in fig. \ref{fig:data_on}, as previously mentioned, requires further investigation in the future.

In conclusion, we have established a microscopic relation between the magnetic susceptibility and the scattering parameter $|S_{11}|$ for cavity-spin systems in general. This framework was applied to investigate the strongly hybridized electron-nuclear spin states across the quantum phase transition in LiHoF$_4$. Employing a re-entrant cavity resonator we achieved both off-resonance and resonant strong coupling and were able to relate the measured $|S_{11}|$ parameters directly to the calculated generalized susceptibility of LiHoF$_4$. Fitting the individual spectra we were able to extract coupling parameter and line width for each of the transitions. The extracted coupling parameters confirmed Boltzmann distribution of the nuclear spin states, while the line widths displayed a non-monotonic dependency as function of transition number, which remains to be explained. Remarkably, with essentially no adjustable parameters both the overall spectra and the extracted coupling parameters are reproduced quite accurately by the established model for  LiHoF$_4$. This demonstrates that cavity-magnon-polariton experiments can be employed as a powerful spectroscopy technique providing detailed quantitative information of the generalized susceptibility of strongly hybridized collective spin systems.

\section*{\label{sec:Exp} Methods}
\subsection*{\label{sec:theo_method} Theoretical method}
To capture the essence of interactions between a cavity field and a magnetic system, it is essential to examine the dynamical magnetic response of the magnetic material. Conversely, from the perspective of the cavity field, there exists a well established theoretical framework, namely input-output formalism, to describe its behaviors. In this section, we first recall the calculation of susceptibilities using linear response theory. Subsequently, starting with the full Hamiltonian of the cavity-spin system, we derive an expression for the scattering parameter $|S_{11}|$ using input-output formalism with the use of explicit spin operators instead of resorting to RWA on the spin-photon interaction. Finally, we establish an explicit link between these two results to provide a simpler yet more effective description for multi-mode magnon-polariton systems in general.

\subsection*{Generalized susceptibility from linear response theory}
\label{linearRT}
The theoretical treatment of the dynamics of a quantum mechanical system subject to an external, time-dependent and perturbative parameter is often well captured by the linear response theory. Let the Hamiltonian of a closed quantum mechanical system be denoted by $\ham_0$. If we introduce a time-dependent perturbation to the system by the environment through an operator $\hat{A}$, then total Hamiltonian can be written as $\ham = \ham_0 + f(t)\mathcal{\hat{A}}$, where $f(t)$ describes the time-dependence of the perturbation. The eigen-states and thus the expectation values of all the physical observables of the original system will vary in response to the perturbation, and these changes can be quantified by a generalized susceptibility. For instance, consider a specific operator $\mathcal{\hat{B}}$. If a complete set of eigenstates of the system is known, the susceptibility can be directly calculated and expressed as follow \cite{Jensen1991}:
\begin{equation}
\label{Eq:gsus}
\chi_{\mathcal{A}\mathcal{B}}(\omega) = \sum\limits_{mn}\frac{\langle m|\hat{\mathcal{A}}|n\rangle\langle n|\hat{\mathcal{B}}|m\rangle}{E_m-E_n-\hbar\omega - i\hbar\gamma_{mn}}(N_m-N_n)
\end{equation}
where $m$ and $n$ index through all the eigenstates and $E_{m,n}$ represent the corresponding eigen-energies. $\w$ is the angular frequency, and $\gamma_{mn}$ is a vanishing quantity that, in real systems, represents the spectral line-width of the transition between eigenstates $|m\rangle$ and $|n\rangle$. $N_{m,n}$ are the occupation factors determined by the partition function and Boltzmann distribution. As there is no restriction on operators $\mathcal{\hat{A}}$ and $\mathcal{\hat{B}}$, one can therefore use Eq.\ref{Eq:gsus}, for example, to calculate AC susceptibilities due to interactions between an applied magnetic field and an ensemble of magnetic moments. In this scenario, $\hat{\mathcal{A}} = \hat{\mathcal{B}} = \sum_{r}\hat{\mu}^{\alpha}_r$, where $\hat{\mu}^\alpha_r$ is the magnetic moment on site $r$ with a spin along the polarization $\hat{\alpha}$.

\subsection*{Mean Field Model of LiHoF$_4$}
All contributions to the net electronic magnetic moment of LiHoF$_4$ originate from the Ho$^{3+}$ ions, which gives an effective spin of $J = 8$ according to Hund's law. The resulting $2J + 1 = 17$-fold degeneracy of the spin degree of rfeedom is lifted due to the presence of the crystal electric field (CEF), producing a ground state doublet situated 0.95 meV (229 GHz) below the first excited state. The spin-spin interaction in LiHoF$_4$, predominately dipolar in nature, induces spontaneous symmetry breaking and ferromagnetically orders the compound along its easy axis at temperatures below 1.53 K \cite{Bitko1996,Ronnow2007}. Furthermore, each Ho nucleus has a spin of $I = 7/2$, and a strong hyperfine interaction local to each Ho site further splits the electronic states of LiHoF$_4$. In addition, the hyperfine interaction also modifies the magnetic phase boundary of LiHoF$_4$ at temperatures below 300 mK and halts the complete softening of the lowest electronic excitation\cite{Ronnow2005}.

LiHoF$_4$ belongs to the space group \textit{I}4$_1$/a, n$^\circ$88 and has a tetrahedral scheelite structure with lattice constants of $a = b = $ 5.164 \AA, and $c = $ 10.78 \AA that is also the easy axis. The full spin Hamiltonian of LiHoF$_4$ has been well established through many complementing experiments and theoretical discussions \cite{Bitko1996,Ronnow2005,Chakraborty2004,Kovacevic2016,Tabei2008}, and can be written in the following terms: $\ham = \ham_{CF} + \ham_Z^J + \ham_Z^I + \ham_{hyp} + \ham_{ex} + \ham_{dip}$. Going from left to right, these terms are, respectively, the crystal field Hamiltonian, the electronic Zeeman interaction, the nuclear Zeeman interaction, the hyperfine interaction, the exchange interaction, and the dipole-dipole interaction. The first four terms are all single-site interactions, whereas the last two are pairwise interactions that grant the system its many-body nature. Details of the crystal field parameters are available in numerous previous reports \cite{Beauvillain1980,Ronnow2007,Babkevich2015}, and for the present study, we adopt those provided in Ref. \cite{Ronnow2007}. It is important to note that the coordinate system used for the CEF parameters in all of the aforementioned reports differ from the crystallographic axes by an $-11^{\circ}$ rotation around the c-axis \cite{Ronnow2007,Babkevich2015}. While this discrepancy has minimal impact on calculations of the electronic spin states, it produces a sizeable effect in the hyperfine level calculations. We therefore rectify it by projecting the crystal field Hamiltonian onto the crystallographic coordinate for consistency, details of which can be found in appendix \ref{appx_CF}.

The transverse component of the electronic Zeeman interaction, $g_L\mu_B\sum_i\vec{J}_i\cdot\vec{B}_\perp$, serves as a source of quantum fluctuations that drives the system to undergo a QPT at 4.95 T at the lowest temperature limit \cite{Ronnow2005,Bitko1996}. The detailed phase diagram has been both experimentally and theoretically investigated as shown in figure \ref{fig:setup}(a). The hyperfine interaction can be explicitly written as $\ham_{hyp} = \sum_iA\vec{I}_i\cdot\vec{J}_i$, where $A$ is the interaction strength, and $\hat{J}_i$ and $\hat{I}_i$ represent, respectively, the electronic and the nuclear spin operators. This interaction mixes the electronic and nuclear spin states, expanding the single-site Hilbert space to (2J+1)$\times$(2I+1) = 136 dimensions ($J = 8$, $I = 7/2$) for LiHoF$_4$. Although the present experiment is carried out at temperatures well below the energy gap (11 K) between the electronic ground state and first excited state, we do not truncate the Hilbert space spanned by the single-ion eigenstates. The value of the hyperfine interaction strength $A$ can be obtained from fitting to experimental data, and for the present report, we again adopt the value reported in \cite{Ronnow2007}. 

\subsection*{\label{sec:Exp_method} Experimental method}
In figure \ref{fig:setup}(c), we present the sample used in our experiment: a thin slab of a single crystal LiHoF$_4$ ($a\times b\times c = 3.0\times0.8\times3.0$ mm, singled out by the red dashed box, is mounted onto a holder made of high purity-copper. The sample is reinforced with a piece of sapphire substrate, singled out by black dashed boxes, on either side of the bigger surface (crystallographic ac-plane). This is intended to improve thermalization and fortify the sample against the large magnetic torque exerted by the transverse magnetic field during experiments. The sample is then installed in a closed 3D two-loop-one-gap re-entrant cavity resonator, seen in figure \ref{fig:setup}(d), that has a diameter of 24 mm and a height of 5 mm. The resonant frequency of our re-entrant cavity can be adjusted by varying the cavity gap size, achieved by changing the inner insert in practice. For the present study, we varied gap size between 200 $\mu$m and 400 $\mu$m, enabling us to alter the frequency of the fundamental mode between $\sim$ 2 GHz and $\sim$ 5 GHz. 

As a re-entrant cavity, which separates the electric and magnetic components of its cavity EM field, our cavity confines the electric field within the gap while the magnetic field fills up the space surrounding it. The field distribution inside the cavity is simulated using finite element method (FEM) by COMSOL$^{\text{\textregistered}}$, the result of which is presented in figure \ref{fig:setup}(d). The color map in the figure illustrates the amplitude of the magnetic component of the cavity EM field for the fundamental mode, while the black arrows indicate the polarization. The solid black rectangles represent the single crystal sample of LiHoF$_4$. In the case of LiHoF$_4$, the strongest response to magnetic probe field comes from its easy axis (c-axis). Therefore, for the present report, the sample is oriented such that its easy axis is parallel to the magnetic polarization of the cavity field. The inhomogeneity of amplitude of the EM field is found by calculations to be less than 2\% across the sample, and the variation in field orientation is found to be about 5\% across the sample. In addition, by using $\rho_s = 4/(5.17\times5.17\times10.75)$ {\AA}$^{-3}$ for the spin density and the physical dimensions of the cavity and the single crystal sample used in the experiment, we computed the filling factor $\eta$.

For the spectroscopic measurement presented in this report, a Vector Network Analyzer (Rohde \& Schwarz$^{\text{\textregistered}}$ ZVL-6) is used to generate the radio frequency signal in the continuous wave (CW) mode at room temperature. The signal is then transmitted to the cavity resonator at milli-kelvins through segments of copper-clad-steel ($>$ 4K) and superconducting NbTi ($\leq$ 4K) semi-rigid coax cables, which are thermalized at every stage of the dilution refrigerator (Oxford Instrument$^{\text{\textregistered}}$ Kelvinox-400). The transmission line is eventually coupled to the cavity resonator via a single-loop-antenna, with an excitation power maintained well within the linear response regime at $\leq -25$ dBm. The impedance matching condition is determined by the relative angle between the antenna's loop plane and the local magnetic polarization of the cavity EM field. Together with the reproducible change in the permeability of the LiHoF$_4$ sample phase transitions, they enable us to reliably reach near-perfect matching conditions at low temperatures by adjusting the antenna's angle between cool-down cycles of the setup. All experimental data are taken in the form of the scattering parameter $|S_{11}|$, measured in a narrow frequency window ($\leq$ 240 MHz) while the external magnetic field is swept at at rate of 20 mT/min using a vertical cryo-magnet up to 17 T. In this report, we demonstrate the validity of our theoretical model by analyzing experimental data from both off-resonance measurement around 4.73 GHz and on-resonance measurement around 3.64 GHz, as indicated by the green dashed horizontal lines in figure \ref{fig:setup}(e).

\section*{\label{sec:ackn} Acknowledgement}
The authors would like to thank Dr. M. M\"{u}ller, Dr. A. Turrini, and Dr. J. Jensen for the helpful discussions. This work was funded by the European Research Council (ERC) under the European Union’s Horizon 2020 research and innovation program projects HERO (Grant No. 810451) and Swiss National Science Foundation (SNSF) grant No. 188648.

\section*{Data availability}
The data that support the findings of this study are available from the corresponding authors upon reasonable request.

\section*{Author contribution}
H.M.R. conceived and co-supervised the project with I.Z.. Y.Y. carried out the experiment with the help of R.G., and P.B. contributed to the design of the experiment. Y.Y. analyzed the experimental results, derived the theoretical results and carried out the simulations. Y.Y. and H.M.R. wrote the manuscript. P.B. contributed parts of the supplementary materials. 

\section*{Additional information}
\textbf{Competing interests}: The authors declare no competing interests.\\

\newpage
\setcounter{figure}{0}
\renewcommand{\thefigure}{S\arabic{figure}}
\subsection*{Cavity field-Spin Interaction}\label{appx_cmp}
The interaction between the cavity photon field and a magnetic sample is in essence Zeeman interaction:
\begin{align}
    \ham^{int}_{cav-spin} &= -\int_{V_s}\vec{\mu}\cdot\vec{B}(\mathbf{r})dr\\
    &=-\sum\limits_c\int_{V_s}\sqrt{\frac{\mu_0\hbar\w_c}{2}}(\hat{a}_c+\hat{a}_c^\dg)\vec{\mu}\cdot\vec{B}'(\mathbf{r})dr
\end{align}
where $\vec{\alpha}$ stands for the magnetic moment and the integration is over the sample volume ($V_s$), and we used the usual quantization rule on cavity photon field \cite{Garrison2008}. For arbitrary shaped cavity, we normalize the field amplitude in the same fashion as in Ref. \cite{Flower2019}, thus the interaction Hamiltonian reads:
\begin{equation}
\begin{split}
    \ham_{cav-spin}^{int} &=-\sum\limits_c\sqrt{\frac{\mu_0\hbar\w_c}{2}}(\hat{a}_c+\hat{a}_c^\dg)\frac{\int_{V_s}\vec{\mu}\cdot\vec{B}'(\mathbf{r})dr}{\sqrt{\int_{V_c}|\vec{B}'(\mathbf{r})|^2dr'}}\\
    &=-\sum\limits_c\sqrt{\frac{\mu_0\hbar\w_c}{2}}(\hat{a}_c+\hat{a}_c^\dg)\int_{V_s}dr\big(\frac{1}{\sqrt{N_s}}\sum\limits_\mathbf{q}\vec{\mu}_q e^{i\mathbf{q}\cdot\mathbf{r}}\big)\cdot\int\frac{d^3\mathbf{k}}{(2\pi)^3}\frac{\vec{B}'(\mathbf{k}) e^{i\mathbf{k}\cdot\mathbf{r}}}{\sqrt{\int_{V_c}|\vec{B}'(\mathbf{r}')|^2dr'}}\\
    &= -\sum\limits_c\sqrt{\frac{\mu_0\hbar\w_c}{2}}\frac{\hat{a}_c+\hat{a}_c^\dg}{\sqrt{\int_{V_c}|\vec{B}'(\mathbf{r}')|^2dr'}}\frac{1}{\sqrt{N_s}}\sum\limits_\mathbf{q}\int\frac{d^3\mathbf{k}}{(2\pi)^3}\vec{\mu}_q\cdot\vec{B}'(\mathbf{k})\int_{V_s} e^{i(\mathbf{q}+\mathbf{k})\cdot\mathbf{r}}dr\\
    &= -\sum\limits_c\sqrt{\frac{\mu_0\hbar\w_c}{2}}\frac{\hat{a}_c+\hat{a}_c^\dg}{\sqrt{\int_{V_c}|\vec{B}'(\mathbf{r}')|^2dr'}}\frac{1}{\sqrt{N_s}}\sum\limits_\mathbf{q}\int\frac{d^3\mathbf{k}}{(2\pi)^3}\vec{\mu}_q\cdot\vec{B}'(\mathbf{k})\frac{N_s}{V_s}\delta(\mathbf{q}+\mathbf{k})\\
    &= -\sum\limits_c\sqrt{\frac{\mu_0\hbar\w_c\rho_s}{2}}(\hat{a}_c+\hat{a}_c^\dg)\sum\limits_\mathbf{q}|\vec{\mu}_q|\sqrt{\frac{[\int_{V_s}|\vec{B}'_\mu(\mathbf{r})|e^{i\mathbf{k}\cdot\mathbf{r}}dr]^2}{V_s\int_{V_c}|\vec{B}'(\mathbf{r}')|^2dr'}}     
\end{split}  
\end{equation}
where in the second line we introduced Fourier transformations $\vec{\mu} = \frac{1}{\sqrt{N_s}}\sum_\mathbf{q}\vec{\mu}_qe^{i\mathbf{q}\cdot\mathbf{r}}$ and $\vec{B}'(\mathbf{r}) = \int\frac{d^3\mathbf{k}}{(2\pi)^2}\vec{B}'(\mathbf{k})\cdot e^{i\mathbf{k}\cdot\mathbf{r}}$, and again the total number of spins in the sample $N_s = V_s \rho_s$. In the last step, $\vec{B}'_\mu(\mathbf{r})$ is the magnetic component of the photon field projected along the sample spin polarization. As we haven't imposed any artificial restrains, the result should be in principle true for any cavity and lattice spin system.

\subsection*{Crystal Field Rotation}\label{appx_CF}
In LiReF$_4$, the $R$ ions are located at a site with local $\bar{4}$ ($S_4$) local symmetry. The crystal field Hamiltonian takes the form,
\begin{equation}
\mathcal{H}_{\rm CEF} = \sum_{l =2,4,6}  B_l^0{\mathbf O}_l^0 +
\sum_{l = 4,6}  B_l^4{\mathbf O}_l^4 + B_l^{-4}{\mathbf O}_l^{-4},
\label{eq:CEF_Ham}
\end{equation}
where Stevens operators ${\mathbf O}_l^{4}$ and ${\mathbf O}_l^{-4}$ are sometimes referred to as ${\mathbf O}_l^4(c)$ and ${\mathbf O}_l^4(s)$, respectively. In a coordinate system where $z$-axis is chosen to be along the four-fold axis, by symmetry we have 7 crystal field parameters $B_l^m$ which are to be determined experimentally. A common approach in spectroscopic measurements such as by Hansen \emph{et al.} [PRB {\textbf 12}, 5315 (1975)] is to chose a coordinate system where the $x$ direction is chosen such that $B_4^{-4} = 0$. Furthermore, one can approximate the $\bar{4}$ point symmetry by $\bar{4}2m$ ($D_{2d}$) symmetry, in which case $B_6^{-4} = 0$. This greatly reduces the number of free parameters in the fit. The Stevens operators that we employ here are defined as,
\begin{align*}
{\mathbf O}_2^0 = {}&3{\mathbf J}_z^2-X{\mathbf I}\\
{\mathbf O}_4^0 = {}&35{\mathbf J}_z^4-(30X-25){\mathbf J}_z^2+(3X^2-6X){\mathbf I}\\
{\mathbf O}_4^4 = {}&({\mathbf J}_+^4+{\mathbf J}_-^4)/2\\
{\mathbf O}_4^{-4} = {}&-{\rm i}({\mathbf J}_+^4-{\mathbf J}_-^4)/2\\
\begin{split}
{\mathbf O}_6^0 = {}&231{\mathbf J}_z^6-(315X-735){\mathbf J}_z^4\\
&+(105X^2-525X+294){\mathbf J}_z^2\\
&+(-5X^3+40X^2-60X){\mathbf I}
\end{split}\\
\begin{split}
{\mathbf O}_6^4 = {}&[(11{\mathbf J}_z^2-X{\mathbf I}-38{\mathbf I})({\mathbf J}_+^4+{\mathbf J}_-^4)\\
&+({\mathbf J}_+^4+{\mathbf J}_-^4)(11{\mathbf J}_z^2-X{\mathbf I}-38{\mathbf I})]/4
\end{split}\\
\begin{split}
{\mathbf O}_6^{-4} = {}&-{\rm i}[(11{\mathbf J}_z^2-X{\mathbf I}-38{\mathbf I})({\mathbf J}_+^4-{\mathbf J}_-^4)\\
&+({\mathbf J}_+^4-{\mathbf J}_-^4)(11{\mathbf J}_z^2-X{\mathbf I}-38{\mathbf I})]/4
\end{split}
\end{align*}
where $X=J(J+1)$ and ${\mathbf I}$ is the identity matrix of size $(2J+1)\times(2J+1)$. One can see from the operators that in the case of $\bar{4}$ point symmetry, there will be imaginary terms in the Hamiltonian. In the case of $\bar{4}2m$ local symmetry, the total Hamiltonian can be made real.

For a general rotation $R(\psi,\mathbf{n})$ by an angle $\psi$ about an axis in the direction of vector $\mathbf{n} = (\sin\theta \cos\phi, \sin\theta\sin\phi,\cos\theta)$, by considering a product of infinitesimal rotations, one finds the unitary transformation operator to be,
\begin{equation}
\mathbf{U}(R(\psi,\mathbf{n})) = \exp(-i m \psi \mathbf{n} \cdot \mathbf{J}).
\end{equation}
This result can be used to determine how the operators $\mathbf{J}$ transform under a finite rotation. We can apply the rotation operator on the Hamiltonian where $\mathcal{H}^R_{\rm CEF} = \mathbf{U}^\dag\mathcal{H}_{\rm CEF}\mathbf{U}$. From the Stevens operators above, we find that ${\mathbf O}_l^0$ operators commute with $\exp(-i m \psi \mathbf{J}_z)$ and the rotation operator has effectively no effect on these terms. However, ${\mathbf{O}}_l^{\pm4}$ terms contain combination of raising and lowering operators ${\mathbf J}_+^4\pm{\mathbf J}_-^4$ resulting in transformation of the operators as $\{{\mathbf O}_6^4\} = \cos(4\psi) [{\mathbf O}_6^4] - \sin(4\psi) [{\mathbf O}_6^{-4}]$, etc. This reflects the more general rule that connects rotations of 3D space and their unitary representatives on the Hilbert space,
\begin{equation}
\mathbf{U}^\dag(\psi,\mathbf{n}) \mathbf{V} \mathbf{U}(\psi,\mathbf{n}) = R(\psi,\mathbf{n})\mathbf{V}.
\end{equation}

An alternative way to tackle the problem for arbitrary rotation of the Stevens operators are done by use of the rotation of Racah operators and the fact that the Stevens operators are linear combinations of Racah operators. The Racah operators namely transform under rotations of the frame of coordinates as the spherical harmonics, whereas the Stevens operators transform as the tesseral harmonics. The Stevens operators, denoted ${\mathbf O}_l^m$, have the disadvantage of not having systematic transformation properties under rotations of the frame of coordinates. Another set of operators, the Racah operators, denoted $\tilde{{\mathbf O}}_{l,m}$, are tensor operators and they therefore have systematic transformation properties. The Racah and Stevens operators are related as,
\begin{eqnarray}
{\mathbf{O}}_l^m &=&
\frac{1}{\kappa_l^m}\left(\frac{2l+1}{8\pi}\right)^{1/2}\left[\tilde{{\mathbf O}}_{l,-m} + (-1)^m\tilde{{\mathbf O}}_{l,m} \right]\\
{\mathbf{O}}_l^0 &=&
\frac{1}{\kappa_l^0}\left(\frac{2l+1}{4\pi}\right)^{1/2}
\tilde{{\mathbf O}}_{l,0}\\
{\mathbf{O}}_l^{-m} &=&
\frac{1}{\kappa_l^m}\left(\frac{2l+1}{8\pi}\right)^{1/2}\left[\tilde{{\mathbf{O}}}_{l,-m} - (-1)^m\tilde{{\mathbf O}}_{l,m} \right]
\end{eqnarray}

The normalising coefficients $\kappa_l^m$ in tesseral harmonics are tabulated elsewhere in Danielsen and Lindg{\aa}rd, Ris{\o} Report No 259 (1972). A comprehensive analysis of transformation properties of Stevens operators for a general rotation $(\phi,\theta)$ of the frame of coordinates is described by Rudowicz [J. Phys. C: Solid State Phys. \textbf{18}, 1415 (1985)] and Mulak and Gajek, \emph{The effective crystal field potential}, Elsevier Science Ltd, Oxford (2000). In the appendix Rudowicz gives a summary of a general rotation of Stevens operators. For the case where $\theta=0$, we find that,
\begin{equation}
\{{\mathbf O}_6^4\} =
\cos(4\phi) [{\mathbf O}_6^4] -
\sin(4\phi) [{\mathbf O}_6^{-4}],
\end{equation}
and similarly,
\begin{equation}
\{{\mathbf O}_4^4\} =
\cos(4\phi) [{\mathbf O}_4^4] -
\sin(4\phi) [{\mathbf O}_4^{-4}],
\end{equation}
where the curly brackets denote a column matrix of Stevens operators in the original axis system and the square brackets denote the operator in the transformed axis system which is rotated by an azimuthal angle $\phi$. The expressions for $\{{\mathbf O}_4^{-4}\}$ are obtained by replacing $\cos(m\phi) \Rightarrow \sin(m\phi)$ and $\sin(m\phi) \Rightarrow -\cos(m\phi)$. The transformation matrices ${\mathbf S}_l(\phi,\theta)$ defined,
\begin{equation}
\{{\mathbf O}_l\} = {\mathbf S}_l(\phi,\theta) [{\mathbf O}_l],
\end{equation}
are real but are not orthogonal -- their inverse matrices are not equal to the transposed ones. 

For the special case given in Eq.~\ref{eq:CEF_Ham}, we find that under a rotation about $z$, the Stevens operators transform as,
\begin{equation}
\left\{
\begin{array}{ll}
{\mathbf O}_l^4\\
{\mathbf O}_l^0\\
{\mathbf O}_l^{-4}
\end{array} 
\right\} =
\begin{pmatrix}
\cos(4\phi) & 0 & -\sin(4\phi)\\
0 & 1 & 0\\
\sin(4\phi) & 0 & \cos(4\phi)
\end{pmatrix}
\left[
\begin{array}{ll}
{\mathbf O}_l^4\\
{\mathbf O}_l^0\\
{\mathbf O}_l^{-4}
\end{array}
\right],
\end{equation}
or alternatively from the view point of crystal field parameters,
\begin{equation}
\left[
\begin{array}{ll}
B_l^4\\
B_l^0\\
B_l^{-4}
\end{array}
\right] = 
\begin{pmatrix}
\cos(4\phi) & 0 & \sin(4\phi)\\
0 & 1 & 0\\
-\sin(4\phi) & 0 & \cos(4\phi)
\end{pmatrix}
\left\{
\begin{array}{ll}
B_l^4\\
B_l^0\\
B_l^{-4}
\end{array}
\right\}.
\end{equation}

\subsection*{Re-entrant Cavity Details}\label{appx_cavity}
\begin{figure*}[ht]
\centering
\includegraphics[width=\textwidth]{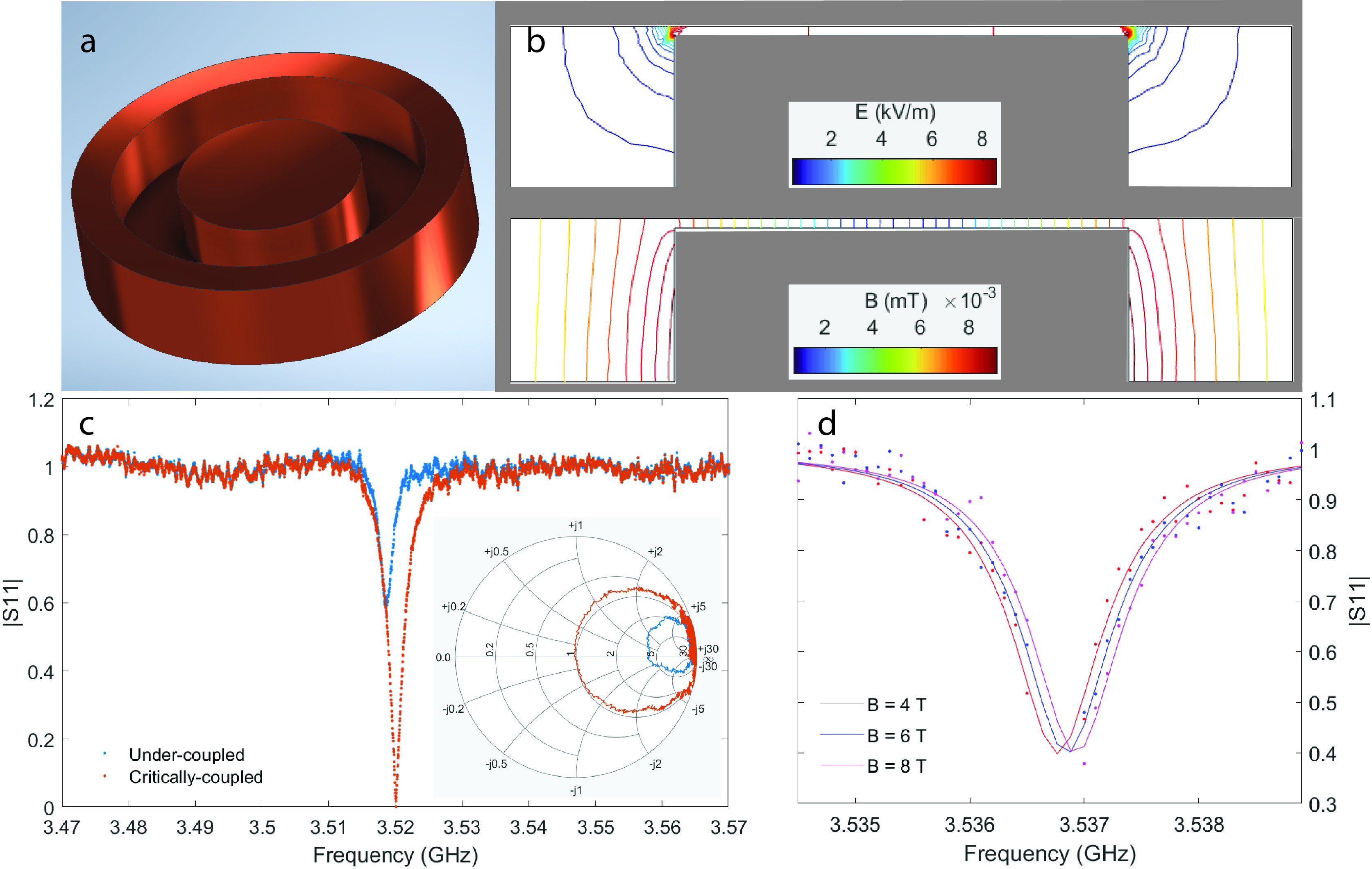}
\caption{\label{sfig:cavity}The re-entrant cavity used in the present work. (a) 3D illustration of the re-entrant cavity, mainly consists of a cylindrical cavity space and an re-entrant post located at the center of the cavity space. (b) Calculated electrical (upper panel) and magnetic (lower panel) field density distribution, shown in contours in the vertical (yz) cross section of the re-entrant cavity. (c) Experimentally measured $|S_{11}|$ at zero magnetic field and room temperature when the cavity is under- (blue) and critically (orange) coupled to the transmission line. Inset: the same measurement on smith chart. (d) Experimentally measured (dots) and fitted (solid lines) of $|S_{11}|$ at 100 mK and various magnetic fields.}
\end{figure*}

The microwave cavity used in the present work is a standard 3D re-entrant cavity with cylindrical symmetry\cite{Hamzah2018}. The cavity is made from high purity copper, whose structure is illustrated in figure S\ref{sfig:cavity}(a). The cavity space is a cylinder of 24 mm in diameter and 5 mm in height, a changeable cylindrical re-entrant post is placed concentric to the cavity space. In order to tune the bare cavity resonant frequency between 2 GHz and 5 GHz, the heights of the re-entrant posts are varied so that the capacitance gap (between the top of the post and the lid of the cavity) ranges from 200 $\mu$m to 600 $\mu$m. The cavity is coupled to the measurement circuit using a single-loop-antenna inserted in the cavity at a location symmetric to the position of the sample.

To characterize the re-entrant cavity, we first use COMSOL Multiphysics to carry out finite element analysis for its eigen-mode. The resultant electrical and magnetic field (density) distribution is shown in, respectively, the upper and lower panels of figure \ref{sfig:cavity}(b) using the vertical cross section of the re-entrant cavity. The shaded area represents the copper walls and the white area represents the cavity space. In order to highlight the capacitance gap, the cross section isn't drawn to scale. We then proceed to measure the scattering parameter $|S_{11}|$ of an empty cavity when it is both critically and over- coupled to the measurement setup. We show an an example of the measurement result in figure \ref{sfig:cavity}(c). Note that the cavity resonant frequency shown in this example is slightly different from the one in the main text, but there is no fundamental difference between the two cavities. We also characterize the field dependence of the bare cavity resonant frequency through measured $|S_{11}|$ at low temperature (100 mK) and various high magnetic fields, the result of which is shown in figure \ref{sfig:cavity}(d). It is clear from fig.\ref{sfig:cavity}(d) that the field dependence of the resonant frequency is very small ($\sim$ 45 kHz/T) comparing to the dramatic shifts due to the LiHoF$_4$ sample in off-resonance measurement (figure \ref{fig:data_off}).

\clearpage
\bibliography{./LiHoF4_cavity_Yikai}

\end{CJK*}
\end{document}